\documentclass[aps,prl,twocolumn,showpacs,psfig,superscriptaddress,longbibliography]{revtex4-1}
\usepackage{times}
\usepackage{graphicx}
\usepackage{float}
\usepackage{latexsym,amsmath,amssymb,bm,euscript}
\usepackage{color}
\usepackage{subfigure}
\usepackage{epstopdf}
\usepackage[colorlinks=true,linkcolor=blue,citecolor=blue]{hyperref}
\usepackage{ulem}% comment it when submit to PRL
\usepackage{enumerate}

\newcommand{\order}[1]{\mathcal{O}{\left(#1\right)}}

\newcommand{\Fig}[1]{Fig.~\ref{#1}}

\def\RP2{\mathbb{RP}^2}

\def\ud{\mathrm{d}}

\graphicspath{{./Figs/}}

\begin{document}

%\title{Logarithmic Rainbow Free Energy on a Topological Manifold}
\title{ {Topological and Geometric Universal Thermodynamics in Conformal Field Theory}}

\author{Hao-Xin Wang}
\email{wanghaoxin@buaa.edu.cn}
\affiliation{Department of Physics, Key Laboratory of Micro-Nano Measurement-Manipulation and Physics (Ministry of Education), Beihang University, Beijing 100191, China}

\author{Lei Chen}
\affiliation{Department of Physics, Key Laboratory of Micro-Nano Measurement-Manipulation and Physics (Ministry of Education), Beihang University, Beijing 100191, China}

\author{Hai Lin}
%\email{hailin@mail.tsinghua.edu.cn}
\affiliation{Yau Mathematical Sciences Center, Tsinghua University, Beijing 100084, China}

\author{Wei Li}
\email{w.li@buaa.edu.cn}
\affiliation{Department of Physics, Key Laboratory of Micro-Nano Measurement-Manipulation and Physics (Ministry of Education), Beihang University, Beijing 100191, China}
\affiliation{International Research Institute of Multidisciplinary Science, Beihang University, Beijing 100191, China}

\begin{abstract}
 {Universal thermal data in conformal field theory (CFT) offer a valuable means for characterizing and classifying criticality. With improved tensor network techniques, we investigate the universal thermodynamics on a nonorientable minimal surface, the crosscapped disk (or real projective plane, $\RP2$). Through a cut-and-sew process, $\RP2$ is topologically equivalent to a cylinder with rainbow and crosscap boundaries. We uncover that the crosscap contributes a fractional topological term $\frac{1}{2} \ln{k}$ related to nonorientable genus, with $k$ a universal constant in two-dimensional CFT, while the rainbow boundary gives rise to a geometric term $\frac{c}{4}  \ln{\beta}$, with $\beta$ the manifold size and $c$ the central charge. We have also obtained analytically the logarithmic rainbow term by CFT calculations, and discuss its connection to the renowned Cardy-Peschel conical singularity.}

\end{abstract}
\date{\today}
\maketitle

\textit{Introduction.---}  Finding universal properties in the manybody system is very important  for understanding critical phenomena \cite{Affleck.I:1986:UniversalTerm, PhysRevLett.56.742, AL-Entropy, Cardy-Peschel-1988}, which constitutes a fascinating and influential topic in diverse fields of physics.  {According} to two-dimensional (2D) conformal field theory (CFT) \cite{CFT-Yellowbook}, universal terms appear  {in the thermodynamics of} critical quantum chains at low temperatures \cite{QPT-Sachdev}, and 2D statistical models  {at the critical temperature}. Among others, logarithmic corrections proportional to ubiquitous central charge $c$ are particularly interesting.  Cardy and Peschel \cite{Cardy-Peschel-1988} showed that free energy contains logarithmic terms due to corners or bulk conical singularities \cite{Vassileva-1991,Costa-Santos-conical-2003,Vernier-Jacobsen-2012,Dubail2013,Bondesan2012,PhysRevE.86.041149, PhysRevE.87.022124, PhysRevE.95.052101,Stephan2014}. This logarithmic term is universal and has profound ramifications in the studies of bipartite fidelity and quantum quenches of (1+1)D models \cite{LBF2011,Dubail2013}, as well as in the corner entanglement entropy of (2+1)D quantum systems \cite{PhysRevLett.97.050404, PhysRevB.90.235106, PhysRevLett.115.021602,LogarithmicZaletel2011}. 

Recently, the universal thermodynamics of 2D CFTs on nonorientable surfaces has been explored, including the Klein bottle ($\mathbb{K}^2$) and M\"obius strip \cite{Tu.h:2017:Klein, Tang.w+:2017:CFT, UniEntropy-2017,Klein-boson}. For diagonal CFT partition functions on the Klein bottle, $F_{\mathcal{K}}= \ln \mathcal{Z}^{\mathcal{K}} = \frac{\pi c}{24 v \beta} L + \ln{k}$, with a universal constant $k=\sum_a d_a/\mathcal{D}$, where $d_a$ is the quantum dimension of the $a$th primary field, and $\mathcal{D}=\sqrt{\sum_a d_a^2}$ is the total quantum dimension. 

 {In this work, we consider the real projective plane ($\RP2$), whose elementary polygon is shown in Fig.~\ref{Fig:CP}(d)}. A specific realization of $\mathbb{RP}^2$ can be achieved by gluing a crosscap with a disk, along the open edge, i.e., a crosscapped disk as shown in \Fig{Fig:CP}(b).  {As a minimal surface, $\mathbb{RP}^2$ is} a building block in constructing other nonorientable surfaces (e.g., $\mathbb{K}^2$).  {Therefore, exploring the possible universal CFT thermodynamics on $\RP2$ is of particular interest.} 

\begin{figure}[tbp]
\includegraphics[angle=0,width=1\linewidth]{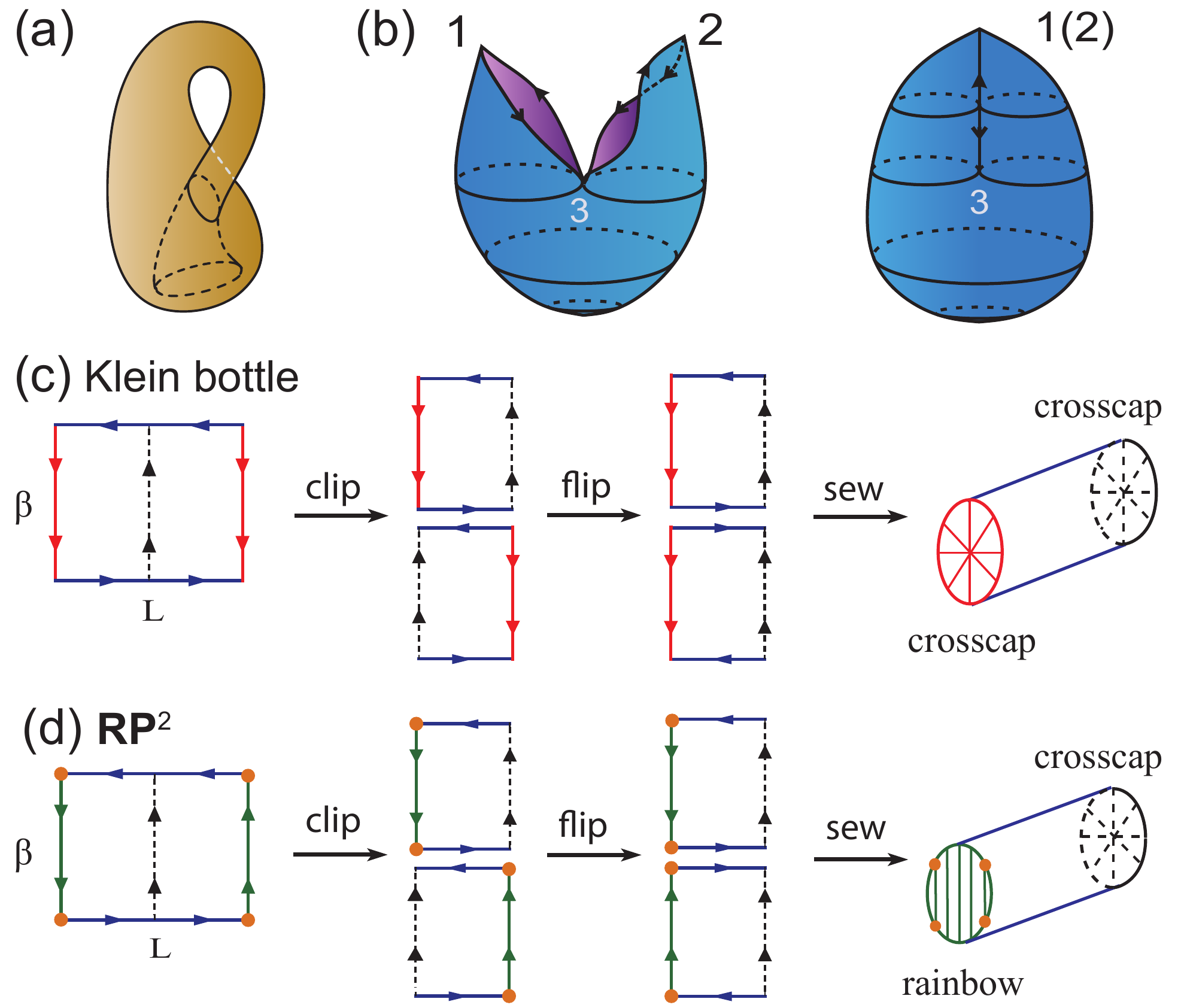}
\caption{Illustration of (a) the Klein bottle ($\mathbb{K}^2$) and  {(b) a crosscapped disk (right) obtained by gluing four lines pairwise with the arrows matched (left).} (c),(d) show the cut-and-sew processes: One clips the elementary polygon vertically along the dashed line, flips the right half horizontally, and properly re-glues the two pieces. $\mathbb{K}^2$ in (c) transforms into a flat cylinder with two crosscap boundaries, while in (d) $\mathbb{RP}^2$ there exist a crosscap and a rainbow.  {The two ``branch" points labeled ``1(2)" and ``3 "in (b), and correspondingly two pairs of orange sites in (d),} play the role of effective conical singularities.}
\label{Fig:CP}
\end{figure}

 {Previous tensor network (TN) methods applied to $\mathbb{K}^2$ (say, in Ref. \onlinecite{UniEntropy-2017}) are not directly applicable to $\mathbb{RP}^2$. Therefore,} we devise here a boundary matrix product state (BMPS) approach, to explore the residual free energy on $\mathbb{RP}^2$.  {This BMPS technique is very efficient and can also improve the accuracy in extracting universal data on other} nonorientable manifolds including $\mathbb{K}^2$ and M\"obius strip, etc.  {The main idea is as follows:} After a cut-and-sew process, $\mathbb{RP}^2$ is transformed into a plain cylinder with special conformal boundaries [Figs.~\ref{Fig:CP}(c) and \ref{Fig:CP}(d)], one crosscap and one so-called rainbow state.  {We find the dominating eigenvector of the transfer matrix through an iterative method, and then extract the universal term by computing its overlap with the crosscap or rainbow boundaries.}

 {With this efficient TN technique removing finite-size effects (in one spatial dimension out of two), as well as CFT analysis, we uncover two universal terms in CFT thermodynamics: the crosscap term $F_{\mathcal{C}}=\frac{1}{2} \ln{k}$, a fractional \textit{topological} Klein bottle entropy due to twist operations; and a \textit{geometric} rainbow term $F_{\mathcal{R}}=\tfrac{c}{4} \ln{\beta}$, as a consequence of the intrinsic ``conical singularity" on $\RP2$,  where $c$ is the central charge and $\beta$ the lattice width (or inverse temperature in quantum cases).}

\textit{Models and TN representations.---} 
We perform TN simulations on the 2D statistical and 1+1D quantum models. The statistical models include the Ising model $H=-\sum_{\langle i,j\rangle} s_is_j$, where $s_i=\pm 1$; the three-state Potts $H=-\sum_{\langle i,j\rangle} \delta_{s_i,s_j}$, with $s_i=0,1,2$, and $\delta$ the Kronecker delta function; and the Blume-Emery-Griffiths (BEG) \cite{BEG1971} model $H = -\sum_{\langle i,j \rangle} s_i s_j+\Delta \sum_i s_i^2$ ($s_i=0, \pm 1$). As shown in \Fig{lattice-and-TN}, we construct the partition-function TN by following either the original lattices where the models are defined [type I, Figs.~\ref{lattice-and-TN}(a) and \ref{lattice-and-TN}(b)], or their ``dual" lattices [type II, Figs.~\ref{lattice-and-TN}(c)--\ref{lattice-and-TN}(e)]. 

 {Type I TN contains vertex tensors $T$ and bond matrices $M$.} For instance, in Fig.~\ref{lattice-and-TN}(a) $T_{s_i,s_j,s_k,s_l} = 1$ (when $s_i=s_j=s_k=s_l$) or 0 (otherwise) is a generalized $\delta$ function, and the matrix $M_{s_m,s_n} = \exp{(- h_{m,n}/T_c)}$ stores the Boltzmann weight [$h_{m,n}$ is the interaction term on the $(m,n)$ bond].  {Hexagonal TN in Fig.~\ref{lattice-and-TN}(b) is constructed similarly, where $T$ is of rank 3.}

 {Type II TN consists of plaquettes/simplex tensors $T$.} In Fig.~\ref{lattice-and-TN}(c), $T$ tensors on (half of) the plaquettes connect each other via $s$ indices and form a square-lattice TN. $T_{s_i,s_j,s_k,s_l}=\exp(-h_{\square_{i,j,k,l}}/T_c)$ and  {$h_{\square_{i,j,k,l}}$ is a plaquette Hamiltonian.} Similarly, we can construct a hexagonal TN representation for the kagome model in Fig.~\ref{lattice-and-TN}(d). For the triangular lattice in \Fig{lattice-and-TN}(e), the dual variables $\sigma_{ij}=s_i s_j$ (Ising) and $e^{2\pi i(s_i-s_j)/3}$ (Potts, see Ref.~\cite{Potts-triangular-Wang}) are introduced on each link $(i,j)$. Correspondingly, the simplex tensor $T_{\sigma_{ij}, \sigma_{jk}, \sigma_{ki}}=\exp(- h_{\triangle_{i,j,k}}/T_c) \delta_{\sigma_{ij} \sigma_{jk} \sigma_{ki}, 1}$, with $h_{\triangle_{i,j,k,}} =- \tfrac{1}{2} (\sigma_{ij} + \sigma_{jk} + \sigma_{ki})$ (Ising) and $-\tfrac{1}{2} [\delta(\sigma_{ij},1) + \delta(\sigma_{jk},1) + \delta(\sigma_{ki},1)]$ (Potts).

 Besides 2D statistical models, critical quantum chains include the transverse-field Ising [$H_{\rm{TFI}} = \sum_{i} (- S_i^x S_{i+1}^x - \tfrac{1}{2} S_i^z)$], Heisenberg XY [$H_{\rm{XY}} = - \sum_{i} (S_i^x S_{i+1}^x + S_i^y S_{i+1}^y)$], and $\mathbb{Z}_3$ quantum Potts [$ H_{\mathrm{Potts}}=-\sum_{i} (\sigma_i \sigma^{\dagger}_{i+1}+\tau_i )+\mathrm{H.c.}$]. The local operators $S^{x,y,z}$ are spin-1/2 operators, and $\sigma_i=\mathrm{diag}(1,\omega,\omega^2)$ with $\omega = e^{2\pi i/3}$, $\tau_i=(e_3,e_1,e_2)$. $e_n$ is a unit column vector with only the $n$th element equal to 1 (others zero) \cite{Parafermion-2015}.  {Given the Hamiltonian $H$, the thermal TN representations of quantum chains can be obtained via the Trotter-Suzuki decomposition of $e^{-\beta H}$, which resemble Fig.~\ref{lattice-and-TN}(c).} 

As follows, we stick to the notation $L$($\beta$) for length(width) in terms of  {TNs, for both square-lattice[Figs.~\ref{lattice-and-TN}(a) and \ref{lattice-and-TN}(c)] and honeycomb-lattice TNs [Figs.~\ref{lattice-and-TN}(b), \ref{lattice-and-TN}(d) and \ref{lattice-and-TN}(e)].} We always assume the thermodynamic limit $L \gg \beta \gg 1$, under which condition the relevant universal terms are well defined. 

\textit{Efficient extraction of universal data.---}  
In previous TN studies \cite{UniEntropy-2017}, given a density operator $\rho$, we evaluate $\mathcal{Z^K}=\rm{Tr} [\Omega \rho(\beta)]$ ($\Omega$ is a spatial reflection operator) to extract universal data on the Klein bottle.The residual term $\ln{k}$ can then be obtained by computing the ratio $k=\frac{\mathcal{Z^K}(2\beta, L/2)}{\mathcal{Z^T}(\beta,L)}$, where $\mathcal{Z^T}(\beta,L)$ is the torus partition function \cite{Tu.h:2017:Klein,Tang.w+:2017:CFT}, 
or by extrapolating $\ln{\mathcal{Z^K}}$ to $L=0$ \cite{UniEntropy-2017}.

\begin{figure}[tbp]
\includegraphics[width=\linewidth]{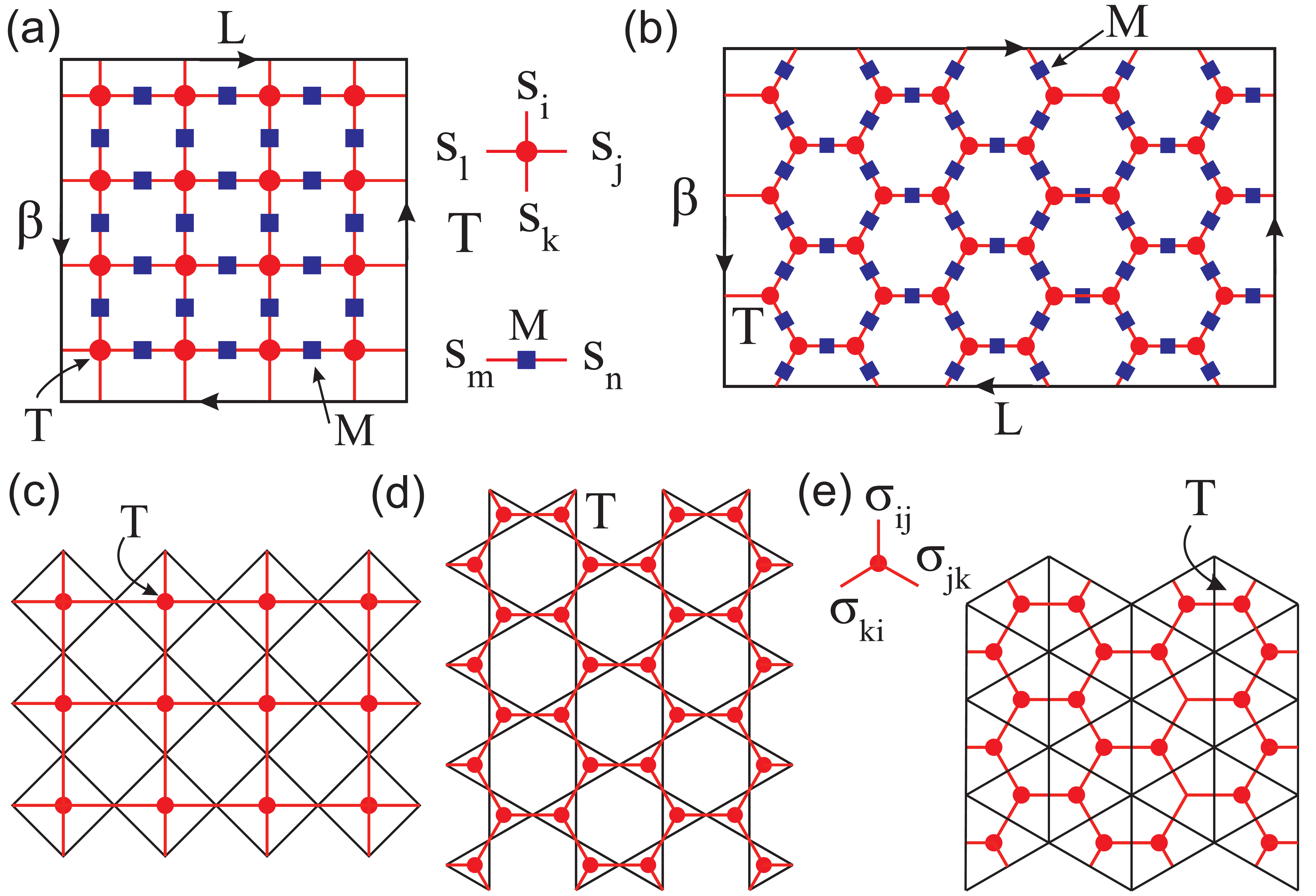}
\caption{Various lattice statistical models [(a,c) square, (b) honeycomb, (d) kagome and (e) triangular] and their corresponding TN representations. (a,b) show TNs defined on the original lattices, and (c,d,e) defined on the ``dual" lattices. To realize the $\RP2$ manifold, we connect square TNs following conventions in (a) and hexagonal TNs like in (b), such that the arrows match.}
 \label{lattice-and-TN}
\end{figure}

\begin{figure}[tbp]
\includegraphics[angle=0,width=1\linewidth]{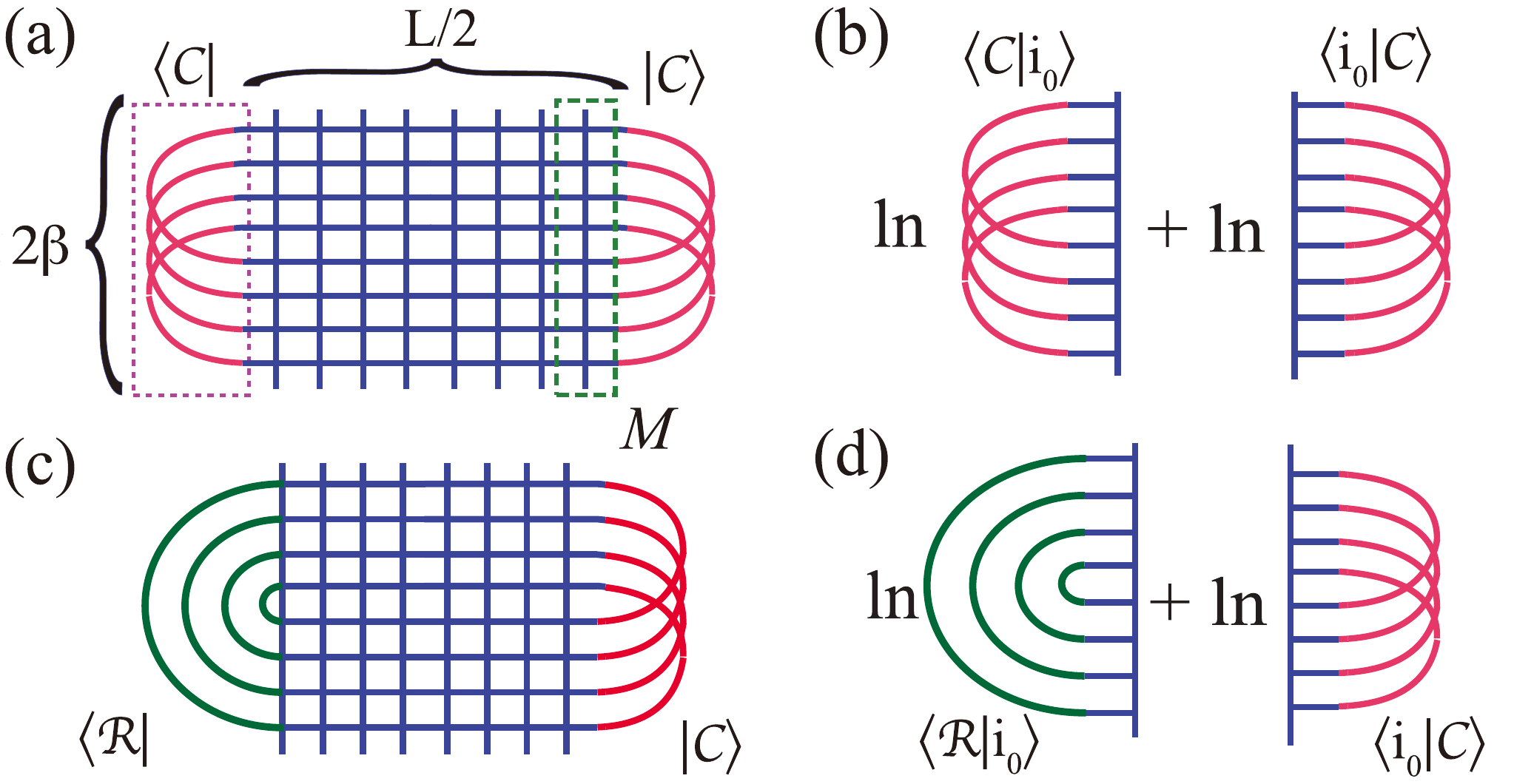}
\caption{(a) TN representation of $\mathbb{K}^2$ partition function consists of columns of transfer matrices ($M$), capped with two crosscaps ($\mathcal{C}$). The dominating eigenstate of $M$ is denoted as $|i_0\rangle$. (b) The Klein term can be computed by summing over logarithm of two overlaps. (c) By substituting one crosscap with a rainbow state $|\mathcal{R} \rangle$, one obtains in (d) the $\mathbb{RP}^2$ universal term. 
}
\label{Fig:Ag}
\end{figure}

However, this scheme is not directly applicable to $\RP2$. Here, we propose a BMPS-based TN technique exploiting the cut-and-sew process in \Fig{Fig:CP}.  Successive projections of the transfer matrix $M$ [see Fig. \ref{Fig:Ag}(a)] to BMPS are performed to determine the (nondegenerate) dominant eigenvector $|i_{0}\rangle$ and then compute the universal data. In practice, $200\!\!-\!\!500$ iterations suffice to converge the bMPS of bond dimension  $D=100\!\!-\!\!500$, offering us results with high precision \cite{Supplementary}.

To be specific, we insert a complete set of orthonormal bases $\{\left| i_{\mu} \right>\}$ into the partition function $\mathcal{Z}(\frac{L}{2}, 2\beta) = \langle \mathcal{B}_L | M^{L/2} | \mathcal{B}_R \rangle$,  {where $\mu$ counts the eigenstates $\left| i_{\mu} \right>$ of the real symmetric transfer matrix $M$. We thus} get \mbox{$\mathcal{Z}=\sum_{\mu} \left< \mathcal{B}_L| i_{\mu} \right> \lambda_{\mu}^{L/2} \left<i_{\mu} | \mathcal{B}_R \right>$}, in which only the dominant eigenvalue $\lambda_0$ and corresponding eigenvectors $\left| i_0 \right>$ survive in the thermodynamic limit, leading to  $\ln{\mathcal{Z}} =\ln[\left<\mathcal{B}_L|i_0\right> \lambda_0^{L/2} \left< i_0|\mathcal{B}_R \right>]=\frac{L}{2} \ln \lambda_0  + F_0$. Clearly, the term $\frac{L}{2} \ln \lambda_0$ corresponds to the bulk free energy.  {Note the transfer-matrix $M$ is exactly the same as that of the Klein bottle (see Fig.~\ref{Fig:Ag}), therefore the universal bulk correction is also $\frac{\pi c}{24 \beta} L$ \cite{UniEntropy-2017}.} Besides, the residual term reads directly as
\begin{equation}
F_0 = \ln{\left<\mathcal{B_L}|i_0\right>} + \ln{\left< i_0 |\mathcal{B_R} \right>}.
\label{Eq:S0}
\end{equation}

 {The cut-and-sew process transforms $\RP2$ into a cylinder with a crosscap $| \mathcal{C} \rangle$ and a rainbow states $|\mathcal{R}\rangle$ on two ends, i.e., $\langle \mathcal B_L|=\langle \mathcal{R} |$ and $|\mathcal B_R\rangle =|\mathcal C\rangle $ [cf. \Fig{Fig:CP}(d)]. Therefore, the residual term $F_0=  F_\mathcal C + F_\mathcal R$, where $F_\mathcal C = \ln \langle i_0|\mathcal C\rangle$ (crosscap term) and $F_\mathcal R = \ln \langle \mathcal R|i_0\rangle$ (rainbow term).}

\textit{Crosscap free energy term.---}
 { $\mathbb{K}^2$ is topologically equivalent to a cylinder with two crosscaps on the boundary, as shown in \Fig{Fig:CP}(c), \ref{Fig:Ag}(a) and \ref{Fig:Ag}(b). Therefore,  $F_{\mathcal{K}}=\ln{\langle \mathcal{C} | i_0 \rangle} + \ln{\langle i_0 |\mathcal{C} \rangle}=\ln{k}$, and it is convenient to see that a fractional term $F_{\mathcal{C}}=\frac{1}{2} \ln{k}$ constitutes an elementary unit of the topological term, associated with a single crosscap boundary $|\mathcal{C} \rangle$.}

Figure~\ref{klein-entropy}(a) shows the Klein and  {crosscap terms} of the {Ising model on} various lattices, which converge to the values $F_{\mathcal{K}}=\ln{(1+\sqrt{2}/2)}$ and $F_{\mathcal{C}}=\frac 1 2 F_{\mathcal{K}}$, respectively. The latter relation is also consistent with the CFT predictions of the $\mathbb{RP}^2$ partition function \cite{Maloney-2016}.   {In particular, one can observe that $F_{\mathcal{C}}$ data converge exponentially fast toward the universal CFT value as $\beta$ increases, regardless of the specific lattice geometries, or even remain identical with $\frac{1}{2} \ln{k}$ up to machine precision [TS case in Fig.~\ref{klein-entropy}(a)].}
%The convergence speed depends on the specific choice of TN representations, e.g., ``TS" representation has much faster convergence (errors $\sim 10^{-15}$, i.e., equals exactly $\ln(1+\frac{\sqrt 2}{2})$ up to machine precision) than that of the ``T" representation on the original square lattice (errors $\sim 10^{-6}$ for width $\beta=8$). 
%\hxx{Remarkably, an analytical solution reveals that even a single row of kagome lattice (i.e., $\beta=1$) presents the universal value exactly \cite{Supplementary}.}\hxc{It's not right now.}

%\hxx{It is natural to attribute one-half of the Klein term (i.e., $\frac{1}{2} \ln{k}$) to each crosscap boundary. In \Fig{klein-entropy}(a), we explicitly show that $F_{\mathcal{C}}=\ln{|\langle \mathcal{C} | i_0 \rangle|} (=\ln{|\langle j_0 | \mathcal{C} \rangle|})$ equals $\frac{1}{2} \ln{k}$ in a very precise way.}

As a useful application, we show that $F_{\mathcal{C}}$ can be employed to accurately determine the critical points, even for challenging models such as the three-state kagome Potts and square-lattice BEG models \cite{Supplementary}. In \Fig{klein-entropy}(b) we show the results for kagome Potts: When $T$ approaches critical temperature $T_c$, $F_{\mathcal{C}}$ converges to $\frac{1}{2}\ln{k}$ ($k={\sqrt{3+\frac{6}{\sqrt{5}}}}$), which otherwise deviates from the universal value. Therefore, from distinct behaviors of the $F_{\mathcal{C}}$ curves [see $|2F_{\mathcal{C}}-\ln{k}|$ in the inset of \Fig{klein-entropy}(b)], we can pinpoint the critical temperature  as $1/T_c\simeq 1.056\ 55(5)$. This value constitutes a  {rather accurate} estimate of $T_c$,  {which is in very good agreement} with $1/T_c = 1.056\ 56(2)$ estimated in Ref.~\onlinecite{Potts-num},  {as well as $1.056\ 560\ 223\ 1(1)$ in  Ref.~\onlinecite{Jacobsen-2016}.} 

 {The crosscap term deeply relates to topology. Based on the above observation for $\RP2$, as well as that in the M\"obius-band case \cite{UniEntropy-2017}, we conjecture that manifolds with a nonorientable genus $\kappa$ (i.e., with $\kappa$ crosscaps) give rise to a universal topological term $\frac\kappa 2 \ln k$.}

\begin{figure}[tbp]
\includegraphics[width=\linewidth]{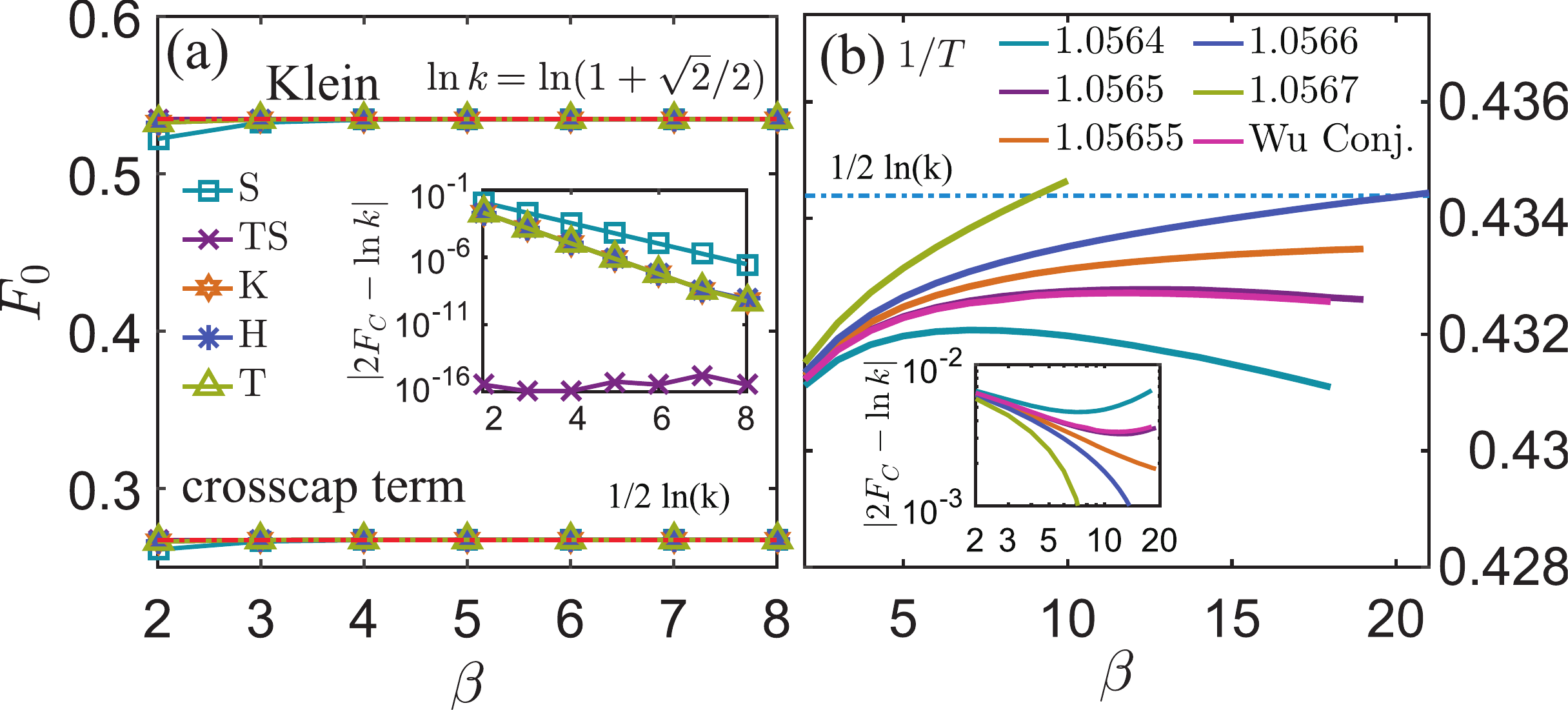}
\caption{(a) The Klein and crosscap terms of Ising models on various lattices.  The crosscap term equals exactly one-half the Klein term. The inset shows the deviation of calculated $F_{\mathcal{K}}$ to the exact value. $D=256$ bond states are retained in the BMPS, i.e, numerically exact. (b) shows $F_{\mathcal{C}}$ of the kagome Potts model at different temperatures, and the inset plots the deviation $|2 F_{\mathcal{C}}-\ln{k}|$.  In this plot, Fig.~\ref{Fig:RainbowEnt}, and Table~\ref{tab:RainbowEnt}, ``S" stands for square lattice, ``TS" for the square lattice with tilted TN representation, ``K" for kagome, ``H" for honeycomb, and ``T" for the triangular lattice. 
\label{klein-entropy}}
\end{figure}

\begin{figure}[tphb]
\includegraphics[width=1\linewidth]{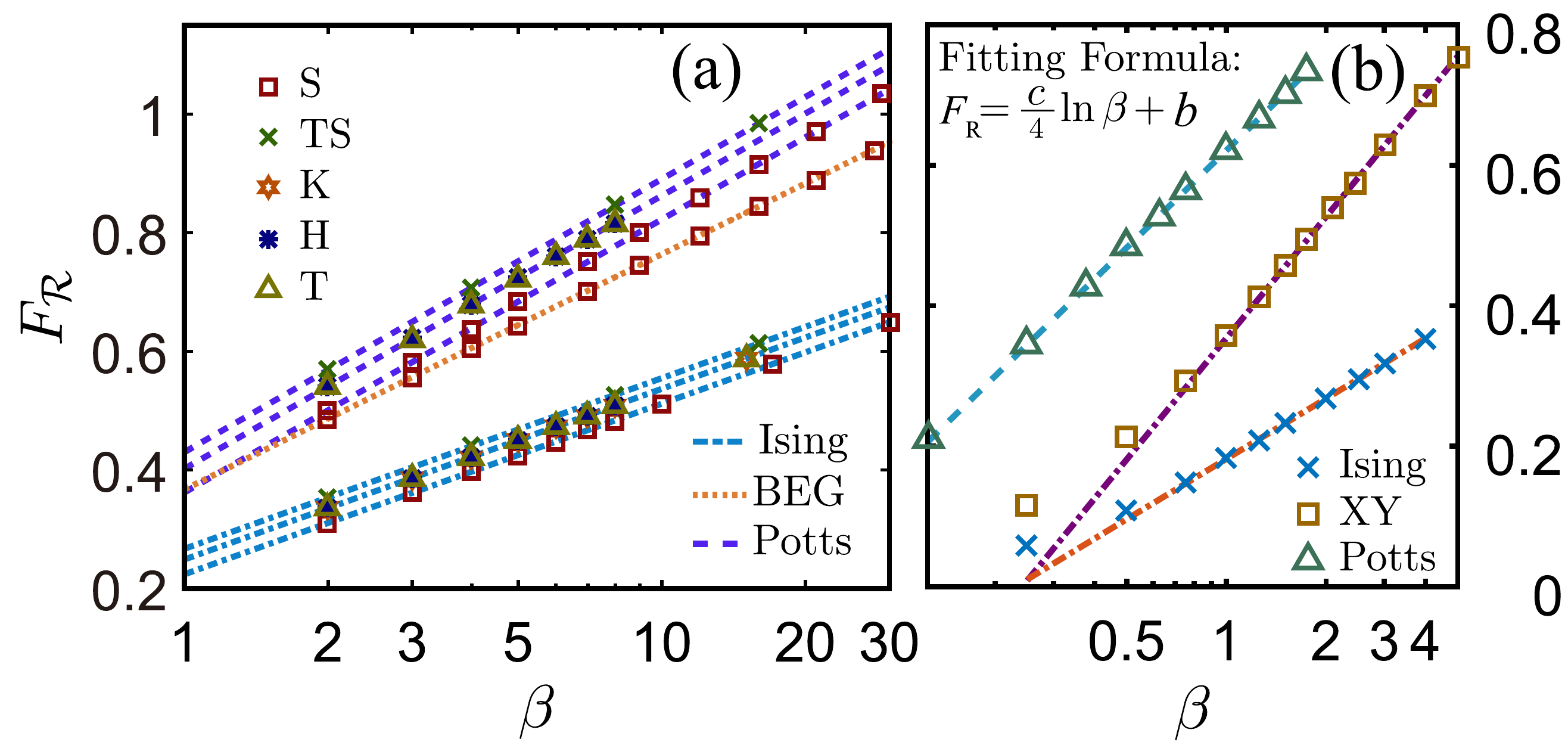} 
\caption{Logarithmic rainbow terms of several critical (a) 2D statistical models and (b) quantum chains.  {For each model, $F_{\mathcal{R}}$ data collapse in three cases, i.e., on kagome, honeycomb and triangular lattices.}}
\label{Fig:RainbowEnt}
\end{figure}

\begin{table*}[htbp]
 % \centering
 \footnotesize
  \caption{Fitted central charge $c$ of 2D lattices (with $z$ the coordination number) and 1D quantum chains (labeled as Q).  {For statistical models, $c$ is fitted with the data of $\beta > 3$; while for quantum models $c$ is fitted in the range $\beta >1$. The numbers in parentheses represent the fitting error bar. } }
    \begin{tabular}{|c|c|c|c|c|c|c|c|c|c|c|c|c|c}
    \hline
    \hline
    Model & \multicolumn{6}{c|}{Ising ($c=0.5$)} & \multicolumn{1}{c}{BEG (0.7)} & \multicolumn{5}{|c|}{Potts (0.8)}   & \multicolumn{1}{c|}{XY (1)}  \\
    \hline
    Lattice & S & {TS} & {H} & {T} & K & Q & S & {S} & TS & H & K & Q & \multicolumn{1}{c|}{Q} \\ \hline
    $T_c$&  \multicolumn{4}{c|}{$\cosh(2/T_c)\cos(\pi/z)=1$ \cite{PhysRev.79.357}} &$\frac{4}{\ln(3+2\sqrt3)}$ \cite{doi:10.1143-ptp-10.2.158}& -&0.609 \footnote{For the BEG model, the tricritical point corresponds to $\Delta=1.966$ at $T_c=0.609$ \cite{PhysRevE.92.022134, PhysRevE.73.036702, PhysRevE.66.026130,PhysRevB.14.4946,Supplementary}.} & \multicolumn{2}{c|}{$\frac{1}{\ln(1+\sqrt3)}$ \cite{RevModPhys.54.235}} & 0.6738 \cite{RevModPhys.54.235} & 0.9465 \footnote{Results in this work, see Fig.\ref{klein-entropy}(b).} & - &  \multicolumn{1}{c|} - \\ \hline
    Slope & 0.1250(2) & 0.1250(2) & 0.1250(1) & 0.1250(1) & 0.1250(1) & 0.124(1) & 0.172(1) & 0.200(1) & 0.200(1) & 0.200(1) & 0.199(1) & 0.1996(5) &   \multicolumn{1}{c|}{0.249(2)} \\ \hline
    fitted $c$ & 0.4998(8) & 0.5000(7) & 0.5000(1) & 0.5000(1) & 0.5000(1) & 0.496(4) & 0.688(4) & 0.800(4) & 0.798(2) & 0.797(2) & 0.804(3) & 0.798(2) &  \multicolumn{1}{c|}{0.996(8)} \\ \hline
    \hline
    \end{tabular}%
  \label{tab:RainbowEnt}%
\end{table*}%

\textit{The rainbow free energy term.---}
Besides the constant crosscap contribution, there exists another logarithmic term in the $\RP2$ free energy due to the rainbow boundary $| \mathcal{R} \rangle$, i.e., 
\begin{equation}
F_{\mathcal{R}} = \tfrac{c}{4 }  \ln{\beta} + b,
\label{Eq:RainbowEnt} 
\end{equation}
where $c$ is the central charge, and $b$ is the non-universal constant.
In \Fig{Fig:RainbowEnt}, we show that, for both 2D statistical models (Ising, Potts, BEG) and 1D critical quantum chains (Ising, Potts and XY),  {the rainbow term scales logarithmically versus $\beta$. The central charge $c$ can be fitted from the slope, and the results are summarized in Table~\ref{tab:RainbowEnt}, where a perfect agreement with CFT is observed. }

Universal logarithmic terms often appear on lattice geometries with corners (such as open strips \cite{Cardy-Peschel-1988,PhysRevE.86.041149,PhysRevE.87.022124}), conical singularities \cite{Costa-Santos-conical-2003}, or even a slit \cite{Dubail2013,PhysRevE.95.052101}. Although $\mathbb{RP}^2$ is a closed manifold without any corners or conical angles, a closer look into the lattice geometry in \Fig{Fig:CP}(d) [and Figs.~\ref{Fig:Ag}(c) and \ref{Fig:Ag}(d)] reveals that there exist two pairs of points which are  {connected twice by lattice bonds (and thus identified twice in the continuous limit)}, forming effective ``conical singularities" on the rainbow boundary. These intrinsic conical points are responsible for the logarithmic term on $\mathbb{RP}^2$.

We provide an intuitive explanation with the multi-scale entanglement renormalization ansatz (MERA) \cite{MERA-2007}, i.e., a holographic view. As shown in  \Fig{Fig:Rainbow}(a), the dominant eigenvector $|i_0\rangle$ of the transfer-matrix is a critical quantum state, 
which has a scale-invariant MERA representation consisting of rank-4 unitary and rank-3 isometry tensors. 
Following the rainbow boundary condition, we fold MERA along two vertical lines and arrive at a double ``MERA" in \Fig{Fig:Rainbow}(b). 
Due to the reflection symmetry, the isometry and unitary tensors in the bulk cancel into identities and do not contribute in $\langle \mathcal{R} | i_0 \rangle$, 
while only the ``corner" tensors (due to folding) contribute to the final trace. 
Note that on the lowest row of MERA the two self-folded corner tensors (colored orange)  connect the four special ``corner" sites indicated in Fig.~\ref{Fig:CP}(d). 
Besides the two corner tensors in the physical layer, 
there exist $\order{\ln{\beta}}$ such ``self-folded" boundary tensors, constrained in the past causal cone [dashed region shown in Fig.~\ref{Fig:Rainbow}(a)] of MERA. 
Each boundary ``impurity" contributes the same factor due to the scale invariance, 
and thus $\langle \mathcal{R} | i_0 \rangle \sim a^{\ln{\beta}}$ (i.e., $\sim \beta^{\gamma}$), giving rise to a logarithmic ``corner" term $F_{\mathcal{R}} \sim \ln{\beta}$.

\begin{figure}[tphb]
\includegraphics[width=0.46\textwidth]{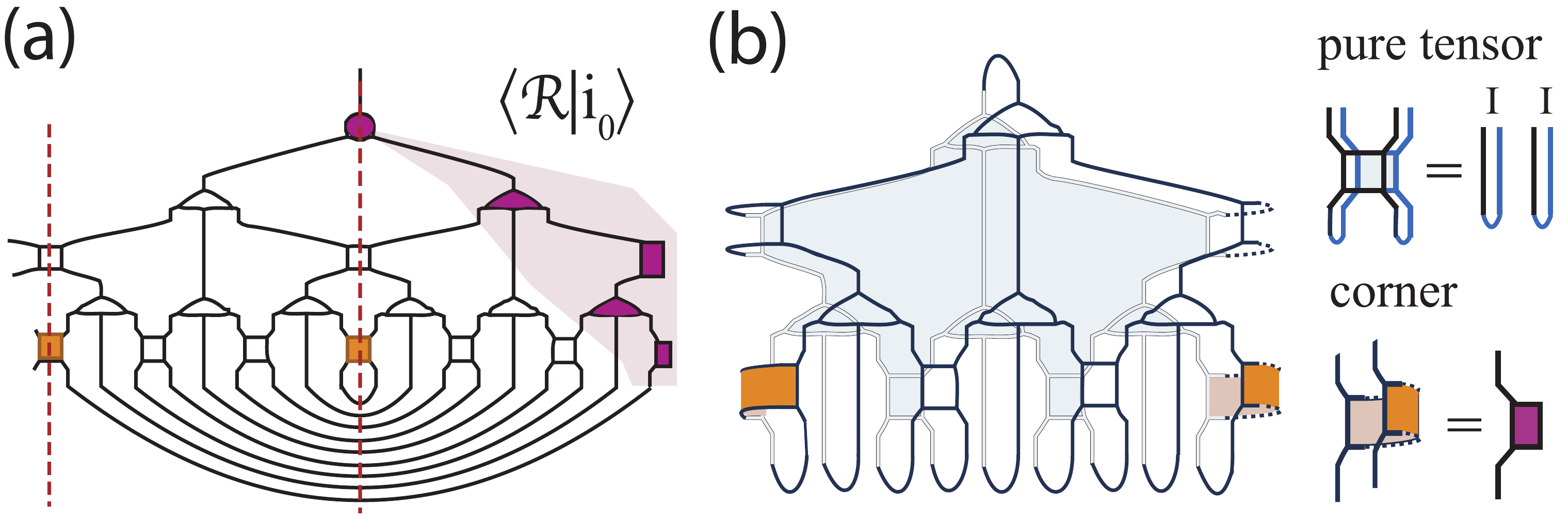}
\caption{(a) Overlap $\langle \mathcal{R} | i_0 \rangle$ of the rainbow state with the scale-invariant MERA can be translated into a folded MERA in (b), where the pure tensors (isometries and unitaries) cancel each other into identities, while the   ``corner" tensors are folded by themselves [right down in (b), coinciding double lines of the original tensor are combined into single indices of the folded one] and located within the boundary causal cone again shown in (a).}
\label{Fig:Rainbow}
\end{figure}
 
\begin{figure}[htbp]
\centering
\includegraphics[width=1\linewidth]{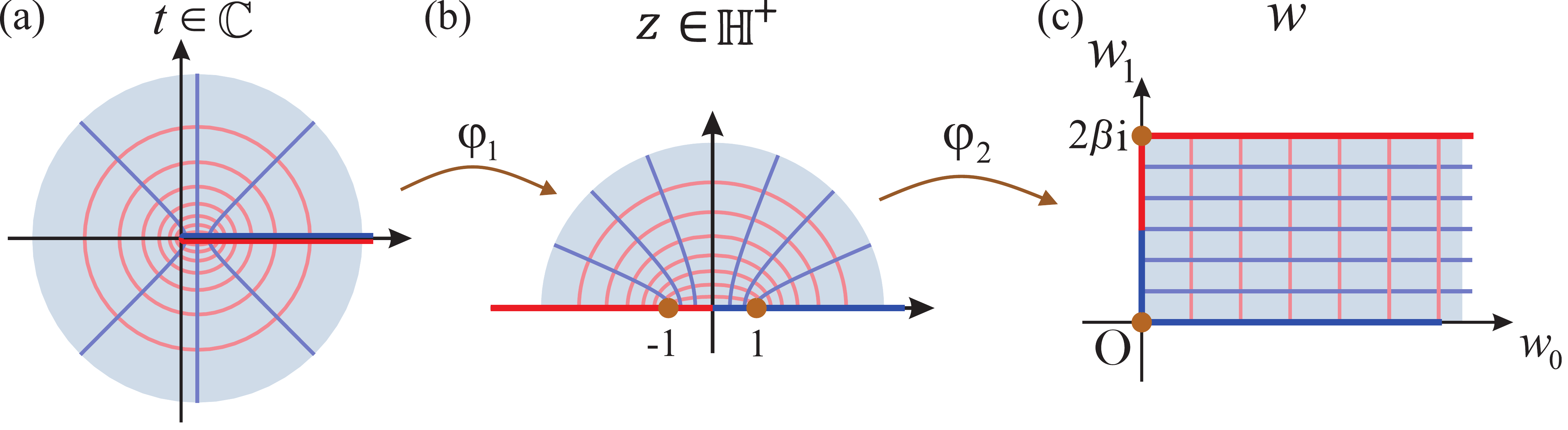}
\caption{ {The conformal transformation $\varphi_1:z=\sqrt{t}$ maps (a) the complex plane to (b) the upper half-plane, which is then mapped onto a semi-infinite rectangle via the Schwarz-Christoffel transformation $\varphi_2:w=\frac{2\beta}{\pi} \mathrm{arcosh}(z)$. Fields defined on the complex plane are assumed to take the same values on the highlighted red and blue lines , giving rise to a rainbow and cylindrical boundary conditions in (c).}\label{map}}
\end{figure}
\textit{CFT analysis of the rainbow term.---}
 {We also analytically derive the rainbow term \cite{Supplementary}, by noting the two successive transformations ($\varphi_1$ and $\varphi_2$ in Fig.~\ref{map}). They map the complex plane to a half-infinite cylinder with a rainbow state on segment [0,$2i\beta$] in Fig.~\ref{map}(c), where $\varphi_2(-1)=2\beta i$ and $\varphi_2(1)=0$. Neglecting the non-universal term and assuming $\langle T(t)\rangle =0$, we obtain $\langle T(w)\rangle = -\frac {c\pi^2} {8(2\beta)^2} [\frac{\sinh^2 ( \pi w/2\beta)}{\cosh^2 (\pi w/2\beta )}+\frac{\cosh^2 (\pi w/2\beta)}{\sinh^2 ( \pi w/2\beta)}]+\frac {c\pi^2}{12 (2\beta)^2}$ according to the transformation of the stress tensor.  As the metric tensor $g_{\mu\nu}$ changes with a tiny variable $\delta g_{\mu\nu}=2\epsilon\delta_{\mu1}\delta_{\nu1}$, the logarithm of the partition function (denoted as $F_0$) varies as  $\delta F_0= \frac 1 2 \int \ud^2 x \sqrt g \delta  g_{\mu\nu } \langle T^{\mu\nu} \rangle = \frac 1 \pi \int_0^{2\beta}\ud w^1 \int_{iw^1}^{L/2+iw^1} \ud w^0 \langle T(w)\rangle \frac{\delta\beta}{\beta} $. We integrate it and arrive at the universal term $F_0 = \frac{\pi c  L} {24 \beta}+\frac c 4 \ln\beta$ .}  {The first term $\frac{\pi c L}{24 \beta}$ also appears in the Klein bottle and M\"obius strip cases \cite{Tu.h:2017:Klein, Tang.w+:2017:CFT, UniEntropy-2017}, as a consequence of the non-orientability for these manifolds. The second $\frac c 4 \ln\beta$ is  the logarithmic rainbow free energy term we have observed numerically.

\begin{table}[htbp]
  \caption{ {Nonorientable universal thermodynamics $F_0$.}}
    \begin{tabular}{cccc}
    \toprule
    Manifolds & Klein bottle & M\"obius strip & $\mathbb{RP}^2$   \\
    \hline
     $\kappa$ & 2 &  1 & 1 \\    
     $\eta$ & $1$  &  $1/2 $ & $1/2$\\ 
     $\mu$ & 0 & 0 & $c/4$\\    
    $\nu$ & $\pi c/24$ 
     &  $\pi c/24$ & $\pi c/24$\\    
    \hline\hline
    \end{tabular}%
  \label{tab:UniEnt}%
\end{table}

\textit{Discussion and summary.---}  The logarithmic rainbow term is geometry dependent and can be related to the renowned Cardy-Peschel conical singularity term
$F_0(\theta)= \frac{c \theta}{24 \pi} [(\frac{2\pi}{\theta})^2-1] \ln{L}$, with $L$ the characteristic system size.
Due to the particular lattice realization of $\RP2$ as in Figs.~\ref{lattice-and-TN}(a) and \ref{lattice-and-TN}(b), there exist two effective $\pi$-angle conical singularities in the TN [see Fig.\ref{Fig:CP}(d)], which in total contribute $2F_0(\pi)=\frac{c}{4} \ln \beta$ (with $\beta \sim L$). Note that the conical angle changes as we alter the specific geometry of the $\RP2$ TN, which introduces a multiplicative geometric factor, following the Cardy-Peschel formula above \cite{Supplementary}.}

 {In Table~\ref{tab:UniEnt}, we briefly summarize the results on nonorientable universal thermodynamics, in the form $F_0=\eta \ln{k} + \mu \ln{\beta} + \nu L/\beta$. Remarkably, the coefficient $\gamma$ is proportional to the nonorientable genus $\kappa$ of the surface, i.e., a topological term; and $\mu$ scales linearly with the central charge $c$, with a geometry-dependent slope. More connections of these universal data to topology and geometry deserve further investigations.}

\textit{Acknowledgments}.--- The authors are indebted to Jin Chen, Xintian Wu, Lei Wang, and Hong-Hao Tu for stimulating discussions. This work was supported by the National Natural Science Foundation of China (Grant No. 11504014).

\title{Supplemental Materials: Topological and Geometric Universal Thermodynamics}

\author{Hao-Xin Wang}
\email{wanghaoxin@buaa.edu.cn}
\affiliation{Department of Physics, Key Laboratory of Micro-Nano Measurement-Manipulation and Physics (Ministry of Education), Beihang University, Beijing 100191, China}

\author{Lei Chen}
\affiliation{Department of Physics, Key Laboratory of Micro-Nano Measurement-Manipulation and Physics (Ministry of Education), Beihang University, Beijing 100191, China}

\author{Hai Lin}
\affiliation{Yau Mathematical Sciences Center, Tsinghua University, Beijing 100084, China}

\author{Wei Li}
\email{w.li@buaa.edu.cn}
\affiliation{Department of Physics, Key Laboratory of Micro-Nano Measurement-Manipulation and Physics (Ministry of Education), Beihang University, Beijing 100191, China}
\affiliation{International Research Institute of Multidisciplinary Science, Beihang University, Beijing 100191, China}

\date{\today}
\maketitle
\widetext

%\newpage
%\mbox{}
%%%%%%%%%% Merge with supplemental materials %%%%%%%%%%
\pagebreak
\begin{center}
\textbf{\large Supplemental Materials: Topological and Geometric Universal Thermodynamics }
\end{center}
%%%%%%%%%% Merge with supplemental materials %%%%%%%%%%
%%%%%%%%%% Prefix a "S" to all equations, figures, tables and reset the counter %%%%%%%%%%
\setcounter{equation}{0}
\setcounter{figure}{0}
\setcounter{table}{0}
%\makeatletter
\renewcommand{\theequation}{S\arabic{equation}}
\renewcommand{\thefigure}{S\arabic{figure}}
\renewcommand{\bibnumfmt}[1]{[S#1]}
\renewcommand{\citenumfont}[1]{S#1}
%%%%%%%%%% Prefix a "S" to all equations, figures, tables and reset the counter %%%%%%%%%%

%\appendix
%\setcounter{figure}{0}    
%\renewcommand\thefigure{\thesection.\arabic{figure}}
%\renewcommand\thefigure{A.\arabic{figure}}

\section{I. Analytical proof for rainbow free energy term}

  {We provide an {analytic calculation of the logarithmic rainbow term from conformal field theory (CFT).} Firstly we introduce two maps:
\begin{enumerate}[(i)]
\item $\varphi_1: \ t \to z =\sqrt t$, where $0\le \arg(z) \le \pi$, $t\in \mathbb C$;
\item $\varphi_2: \ z \to w =\frac{2\beta}{\pi}  \mathrm{arcosh} z$.
\end{enumerate}
}
\begin{figure}[htbp]
\centering
\includegraphics[width=0.8\linewidth]{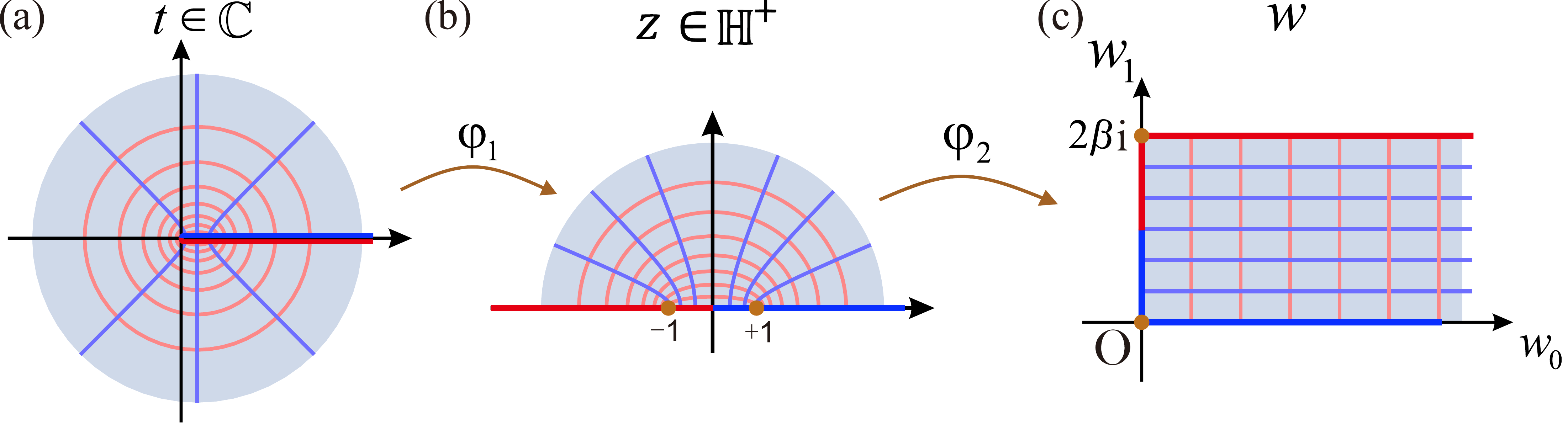}
\caption{{Conformal mappings construct a semi-infinite cylinder geometry with one rainbow boundary.} \label{map}}
\end{figure}

  {$\varphi_1$ maps $\mathbb C$($t$--plane) to $\mathbb H^+$($z$--plane), and {identifies} the negative half real axis with the positive half in $z$--plane, while $\varphi_2$ maps $\mathbb H^+$(in $z$--plane) to a {half-infinite rectangle} $\{w: \Re w \ge  0, 2\beta \ge \Im w \ge 0 \}$ (in $w$--plane). $\varphi_2(-1)=2i\beta$, $\varphi_2(1)=0 $.
The rectangle has reflection symmetry about the line $\{\Im w=\beta \}$, and the {identification in} the real axis on $z$--plane transforms to the {identification} in the reflection symmetric boundaries on the $w-$ plane. Geometrically, {it corresponds to} a {half-infinite} cylinder with  a rainbow boundary, which is along the segment [0,$2i\beta$] on imaginary axis in the rightmost plot of Fig.~\ref{map}.

We consider the transformation of the stress tensor
\begin{equation}
\langle T(z)\rangle = (\frac{\ud t}{\ud z})^{2}\langle T(t)\rangle+\frac{c}{12} \{t;z\},
\end{equation}
where $\{t; z\}= \frac{\ud^3 t/\ud z^3}{\ud t/\ud z} -\frac{3}{2}(\frac{\ud^2 z/\ud t^2}{\ud z/\ud t})^2$ is the Schwarzian derivative. We find that $\{t; z\}= -3/2z^2$,   {which leads to} $\langle T(z)\rangle=-\frac {c}{8z^2}$, by assuming $\langle T(t)\rangle =0$.

Similarly, one can get 
\begin{equation}\label{Tw}
\langle T(w)\rangle = -\frac {c\pi^2} {8(2\beta)^2} [\frac{\sinh^2 ( \pi w/2\beta)}{\cosh^2 (\pi w/2\beta )}+\frac{\cosh^2 (\pi w/2\beta)}{\sinh^2 ( \pi w/2\beta)}]+\frac {c\pi^2}{12 (2\beta)^2 }
\end{equation}
after the conformal map $\varphi_2$.

The universal free energy term $F_0=\ln{\mathcal{Z}_{\rm{CFT}}}$, defined as the   {logarithm} of CFT partition function, varies in the following way as the metric tensor changes by $\delta  g_{\mu\nu }$:
\begin{equation}
\delta F_0 = \frac 1 2 \int \ud^2 x \sqrt g \delta  g_{\mu\nu } \langle T^{\mu\nu} \rangle  
\end{equation}

For such a cylinder with a rainbow boundary, we {introduce} an infinitesimal scaling of the circumference:
$2\beta \to (1+\epsilon)(2\beta)$, {namely,} $\delta \beta = \epsilon \beta$ {with a small $\epsilon$}. This is realized by applying a coordinate transformation $w^1\to (1+\epsilon) w^1$, {and note that $w^1$ is the imaginary part of $w$, i.e., $w=w^0+iw^1$. This transformation leads to the change in the metric tensor $\delta g_{\mu\nu}=2\epsilon \delta_{\mu 1}\delta_{\nu 1}$.} 

According to  
\begin{equation}
   \langle T^{11}\rangle=- \langle T^{00}\rangle=-(T_{ww}+T_{\bar w\bar w})  = \frac 1 \pi  \langle T(w)\rangle, 
\end{equation}
we obtain the variation in $F_0$ as
\begin{equation}\label{F}
\delta F_0=\frac 1 \pi \int \ud w^0\ud w^1 \langle T(w)\rangle \frac{\delta\beta}{\beta}.
\end{equation}
We substitute Eq. (\ref{Tw}) into Eq.~(\ref{F}), and perform the integration as follows.
Firstly, $I$ is introduced as
\begin{equation}
\begin{aligned}
 I &=  \int_0^{2\beta} \ud w^1 \int_{i w^1}^{L/2+i w^1}  \ud w^0 \langle T(w)\rangle \\
  &=  \int_0^{\pi} \ud  w^1 \int_{i w^1}^{\pi L/4\beta +i w^1} \ud w^0 [-\frac c 8 (\frac {\sinh^2  w}{\cosh^2 w }+\frac{\cosh^2 w}{\sinh^2  w}) +\frac{c}{12}],
\end{aligned}
\end{equation}
which is manifestly related to $F_0$ as $L \to \infty $. Exploiting the equation
\begin{equation}
\frac{\ud }{\ud w} (2w-\tanh w -\frac 1{\tanh w} )=\frac{\sinh^2  w}{\cosh^2 w }+\frac{\cosh^2 w}{\sinh^2  w},
\end{equation}
one obtains 
\begin{equation}
I= \int_0^{\pi} \ud  w^1 \{-\frac {c\pi L} {6\cdot 4\beta}  -\frac c 8[i \tan w^1+(i \tan w^1)^{-1} -\tanh (\frac{\pi L}{4\beta} +i w^1) - \tanh^{-1}(\frac{\pi L}{4\beta} +i w^1)]\}
\end{equation}

The integrals of $ \tan w^1$ and $( \tan w^1)^{-1}$ terms cancel each other, and $-\frac {c\pi L} {6\cdot 4\beta}$ term contributes $\frac {\pi c  L} {24 \beta}$ to the resulting free energy term.
%if you integral it and substitute it into Eq. (\ref{F}).
Since $-\tanh (\frac{\pi L}{4\beta} +i w^1) - [\tanh (\frac{\pi L}{4\beta} +i w^1)]^{-1}$ converges to -2 as $L/\beta \to \infty$, they contribute a term $(c/4) \ln\beta$ in the final result, through Eq.~(\ref{F}). All in all, the universal free energy term turns out to be
\begin{equation}
F_0=\frac{\pi c} {24 \beta} L+\frac c 4 \ln\beta,
\end{equation}
which is a sum of the non-orientable correction $\frac{\pi c} {24 \beta} L$ and the logarithmic rainbow term $\frac c 4 \ln\beta$. The former constitutes a bulk correction and affects the slope of low-temperature linear specific heat, in quantum chains.}

%In this calculation, it's not easy to see the relationship between rainbow free energy with corners or conical singularities contribution.
% The poles contribute nothing. It maybe promise new physics.

  {The rainbow term can also be understood from CFT by considering a free boson model. 
A prominent example is the Heisenberg XY spin chain simulated in the main text ($c=1$), whose continuum limit is a CFT modeled by a free boson
with action $S=\frac{1}{4\pi }\int dzd{\bar{z}~\partial \phi }{\bar{\partial}%
}{\bar{\phi}}${, where }${\phi }$ is the boson field. The logarithmic term $F_{\mathcal{R}}$ is due to the effective ``corners", i.e., branch points at $z=0$ identified twice on the rainbow boundary. For $n=c$ copies of free bosons ${\phi }_{i}$, the stress tensor $T(z)=\sum_{i}{\partial \phi }_{i}{\partial \phi }_{i}$ is of scaling dimension two, thus there exists a second-order pole at the $z=0$ branch locus, resulting in the one-point function $\left\langle T(z)\right\rangle = -\frac{c}{8}(1/z^{2})$.
Its variation with respect to $\beta $ is $\delta F_{\mathcal{R}}/\delta \beta =-\frac{%
\delta }{\delta \beta }\left( \frac{1}{2\pi }\int d^{2}z^{\prime
}\left\langle T(z^{\prime })\right\rangle \right) $, and in the
thermodynamic limit, namely large $L$ and $x_{2}/x_{1}\ll 1~$limit [$x_{1}$($%
x_{2}$) is the continuous length(width) integral variable], this is $\delta
F_{\mathcal{R}}/\delta \beta =\frac{\delta }{\delta \beta }\left( \frac{c}{4}%
\int dx_{2}(1/x_{2})\right) =\frac{c}{4\beta }$, i.e., $F_{\mathcal{R}}\sim 
\frac{c}{4}\ln \beta $. }

  {To conclude, here we provide an analytical calculation of the universal free energy term $F_0$. Through two conformal maps which transform the complex plane into half-infinite cylinder with rainbow boundary, we obtain two terms in the final results. They represent, for quantum chains, the bulk free energy correction and rainbow logarithmic entropy, respectively. A simplified argument for the rainbow correction, in free boson field case, is also given.}

\section{II. Boundary matrix product state approach}
\label{SM:BMPS}
In this work, we use power iteration method to determine the eigenvectors of the transfer matrix in tensor networks (TNs), by exploiting the boundary matrix product states (bMPS). We also introduce the TN technique for an efficient extraction of universal data from the obtained bMPS. 

\subsection{A. Determination of dominant MPS by power iterations}
Generally, given an operator $A$, the power iteration algorithm makes use of the recurrence relation
\begin{equation}\label{iteration}
 b_{k+1}=\frac{A \cdot b_k}{||A \cdot b_k||},
\end{equation}
where the vectors $\{b_k\}$, starting with a random vector $b_0$, constitute a converging series. At each iteration, the vector $b_k$ is multiplied by the operator A and then normalized to get the next vector $b_{k+1}$. Power method is typically robust and efficient as a dominant eigenvalue/vector solver. Under the circumstance that $A$ has a unique dominant eigenvalue (with largest magnitude) and the starting vector $b_0$ has a nonzero component parallel to related dominant eigenvector, the power method converges the series \{$b_k$\} right to that eigenvector.

To evaluate the partition functions of quantum chains or 2D statistical models, the power method is combined with the bMPS technique.
We determine the dominant bMPS $b_k$ of a transfer matrix (in the form of matrix product operator, MPO) in the TN by multiplying the MPO $A$ to the MPS $\{b_k\}$ and normalizing it, until the latter converges to the dominant eigenvector (also in the form of bMPS). 

\begin{figure}[htbp]
\includegraphics[width=.8\textwidth]{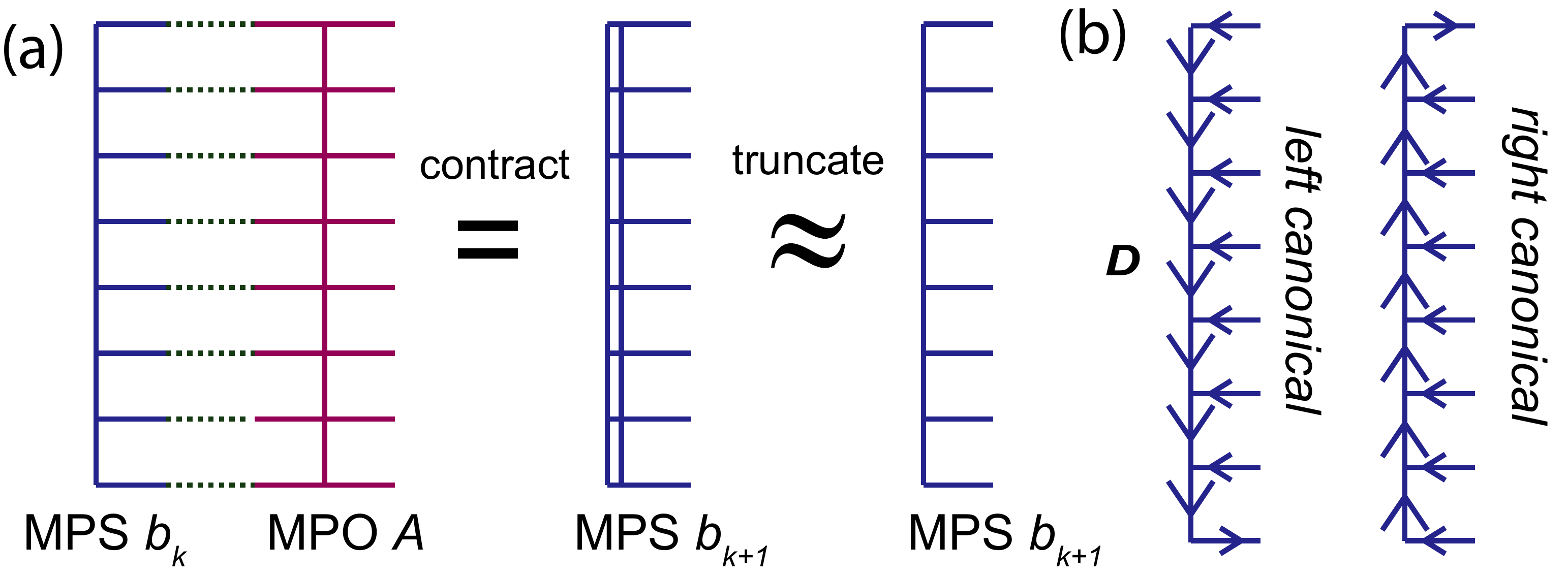}
\caption{(a) Project MPO $A$ to the MPS $b_k$ and obtain a fat MPS $b_{k+1}$, which is then truncated into a MPS with bond dimension $D$. (b) illustrates the forward and backward sweeps in order to guarantee respectively the left- and right-canonical form of MPS, as well as an (globally) optimal truncation process.}
\label{Fig:BMPS}
\end{figure}

Since the Hilbert space grows exponentially with system size, after each iteration we truncate bMPS to a fixed bond dimension, making this procedure sustainable. To be specific, in a single step of multiplying MPO $A$ to MPS $b_k$ [\Fig{Fig:BMPS}(a)], one gets $b_{k+1}$ with a ``fat" MPS, i.e., with an enlarged bond dimension. 
Although the initial MPO $A$ is with periodic boundary condition along the vertical direction [see, e.g., Figs.~\ref{FigS5}(c,d)], we fold the ``long-range" PBC bond in the bulk and use exclusively MPO and MPS with open boundary condition (OBC), for later convenience of canonicalization and truncation processes.

As depicted in \Fig{Fig:BMPS}(b), we introduce a forward and backward canonicalization procedure for the efficient compression, by utilizing the Schmidt decomposition (in practice singular value decomposition, SVD) on each bond bipartition. During the forward canonicalization [left, \Fig{Fig:BMPS}(b)], the MPS is gauged into a left canonical form (see the arrow flow); then we perform the truncation procedure, i.e., keep only the largest $D$ singular values and their corresponding bond basis in the spectra, during the sweep backwards [right, \Fig{Fig:BMPS}(b)].

\subsection{B. Tensor network calculation of the crosscap and rainbow overlaps}%

\begin{figure}[htbp]
\includegraphics[width=\textwidth]{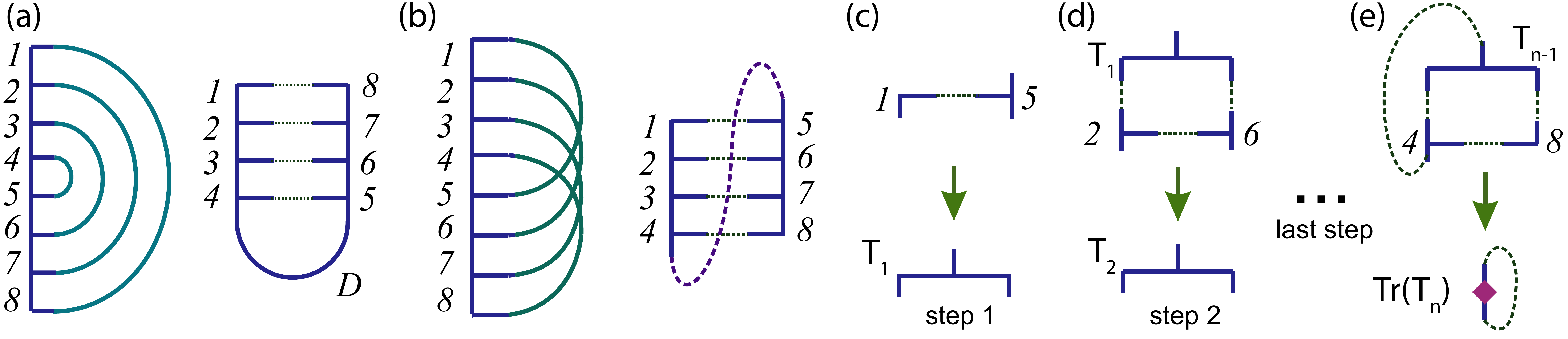}
\caption{Overlap between bMPS and (a) the rainbow state and (b) the crosscap state. (c,d) show the main steps to contract bMPS (of bond dimension $D$) with the crosscap boundary, which turns out to be special inner-product-like contraction procedure in (c,d) with complexity $\mathcal{O}(D^4)$. The last step (e) finishes the contraction by summing over all bond indices, and returns the trace.}\label{SM:overlap}
\end{figure}%

Once the bMPS representation of $| i_0 \rangle$ is obtained, we need to contract the bMPS with corresponding boundary state $| \mathcal{C} \rangle$ (crosscap) or $| \mathcal{R} \rangle$ (rainbow) in the last step to extract universal terms. These overlap calculations can be converted to tensor contractions (traces) shown in \Fig{SM:overlap}. 

We start with the rainbow boundary overlap $\langle i_0 | \mathcal{R} \rangle$ in Fig.~\ref{SM:overlap}(a), which can be conveniently simulated. One folds the bMPS from the middle, and then contract the tensors, just like performing an ordinary MPS inner product, with conventional $\mathcal{O}(D^3)$ complexity ($D$ the bond dimension of bMPS). 

The crosscap boundary overlap $\langle i_0 | \mathcal{C} \rangle$ takes slightly higher computational cost to perform the contraction. As shown in Fig. \ref{SM:overlap}(b), one needs to compute a special inner product of upper half of bMPS with the rest lower half. One starts from the top [see Fig.~\ref{SM:overlap}(c)], prepare a rank-3 tensor $T_1$ and contract successively with the transfer matrix [Fig.~\ref{SM:overlap}(d)], with complexity $\mathcal{O}(D^4)$. After $n$ iterations ($n$ proportional to system width $\beta$), one arrives at a single tensor $T_n$ in Fig.~\ref{SM:overlap}(e) and can compute the trace $\langle i_0 | \mathcal{C} \rangle = \mathrm{Tr}(T_n)$ at ease.

\section{III. Determination of tricritical point of BEG model}%

\begin{figure}[htbp]
\includegraphics[width=0.55\textwidth]{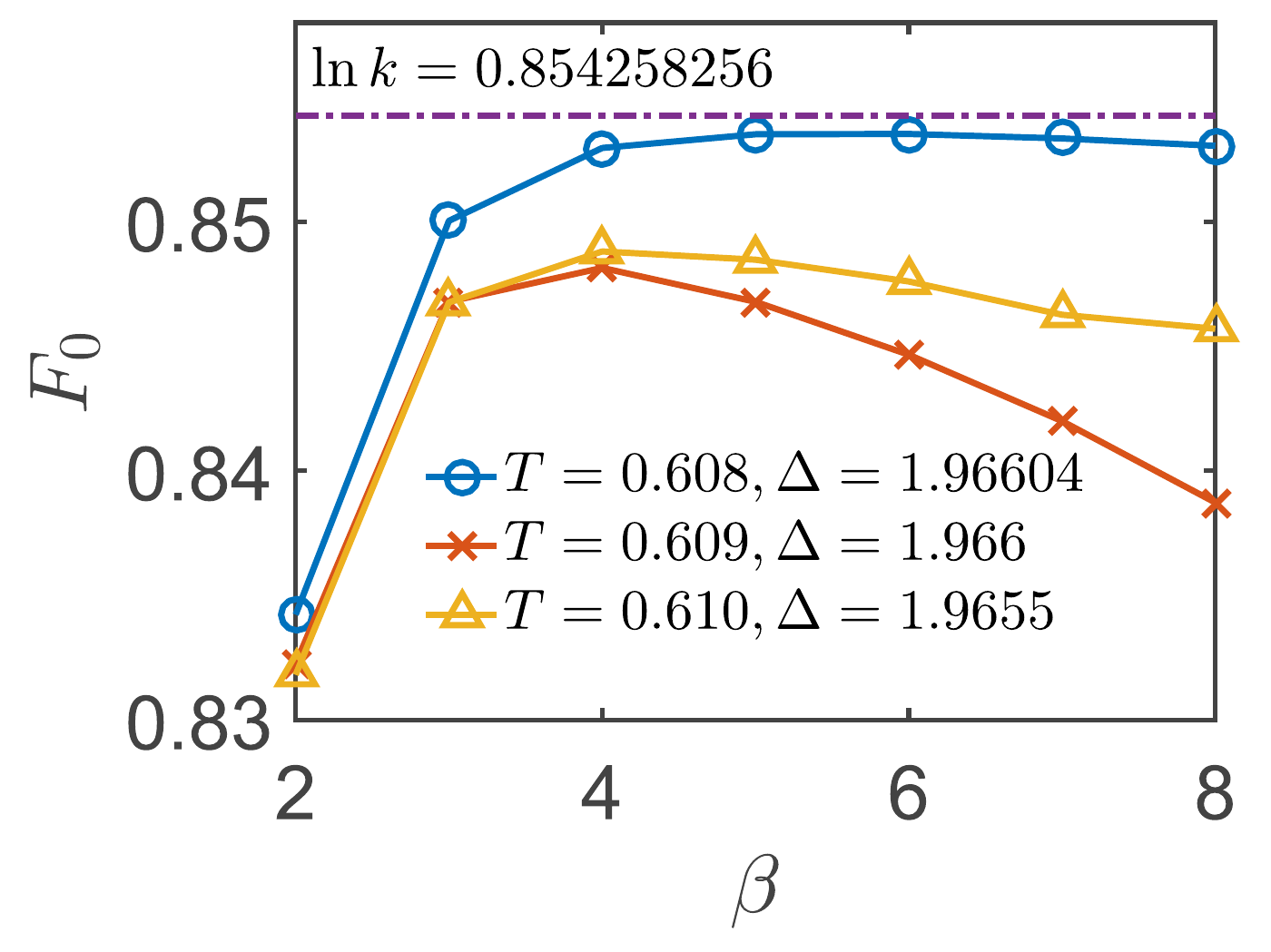}
\caption{Klein free energy results of BEG model, at different temperatures $T$ and spin anisotropy parameters $\Delta$. The parameters $(T, \Delta)$ are previous estimates of tricritical point, taken from Refs. \onlinecite{PhysRevE.92.022134S, PhysRevE.73.036702S, PhysRevE.66.026130S}, in corresponding order. 
\label{SM:BEG}}
\end{figure}
%,PhysRevB.33.1717,1402-4896-90-8-085206,1751-8121-41-40-405004,PhysRevLett.63.660
Recently, there has been considerable interest \cite{PhysRevE.92.022134S,PhysRevE.73.036702S,PhysRevE.66.026130S} in studying the Blume-Emery-Griffiths (BEG) model, both by numerical and analytical methods.
The BEG model is first proposed \cite{BEG1971S} to describe mixtures of liquid He$^3$ and He$^4$, where there exists a superfluid phase, as well as $\lambda$- and first-order transitions, in the abundant He$^4$ phase diagram. 
The Hamiltonian of BEG model reads
\[ E = -J\sum_{<i,j>} s_is_j-K\sum_{<i,j>}s_i^2s_j^2+\Delta \sum_is_i^2, \]
where $s_k=0, \pm 1$ is the spin variable. In Ref.~\onlinecite{PhysRevB.14.4946S},
Berker and Wortis suggested an amenable phase diagram using position-space renormalization group method. It contains three phases: a ferromagnetic phase and two paramagnetic phases (denoted by para$_{\pm}$), characterized by magnetization $M$ and quadrupolar order parameter $Q$. Remarkably, there exists a tricritical line in the phase diagram where three phases coexist, and gives rise to an emergent supersymmetric conformal symmetry.

To the best of our knowledge, no analytical results have been obtained on the precise location of tricritical points of the square-lattice BEG model. Numerical calculations (with $K=0, J=1$) have been performed \cite{PhysRevE.92.022134S, PhysRevE.73.036702S, PhysRevE.66.026130S}, and we here compare these results by calculating the Klein free energy $F_{\mathcal{K}}$ in Fig.~\ref{SM:BEG}. 

We find that $F_{\mathcal{K}}$ can be used to pinpoint the tricriticality in a very sensitive and precise way, and it turns out that the estimate $T=0.608, \Delta=1.96604(1)$ in Ref. \onlinecite{PhysRevE.92.022134S} produces the most accurate $F_{\mathcal{K}}=\ln{k}$, i.e., it is in nice agreement with the CFT prediction of 0.854258 \cite{UniEntropy-2017S}. Therefore, parameter $T=0.608, \Delta=1.96604(1)$ was also adopted in the main text to calculate the rainbow free energy [see Fig. 5 and Tab. I].

%As mentioned in Ref.[DOI: 10.1103/PhysRevB.68.224423], for a conical singularity with spanned angle $\theta$, the contribution to the free energy is 
%\begin{equation}
%F_\theta = \frac {c\theta}{24\pi}[1-(\frac{2\pi}{\theta})^2] \ln L
%\end{equation}
%For a rainbow boundary on square lattice model, $\theta=\pi$ and  $F_\theta = -c/8 \ln L$. Since there are two conical singularities, $F_\mathcal R=-c/4 \ln L$. %

%Similarly, if we regard the conical singularities angles of $C_3$ lattice as $2\pi/3$ and $4\pi/3$, 
%\begin{equation}
%F_\mathcal R = -\frac{c}{4}\frac{1}{6/7}\ln L
%\end{equation}
%$6/7\simeq 0.857142857142$ and $\sin(2\pi/3)=0.866025403784438646$.

\section{IV. Universal term with conical angles other than $\pi$}

  {In \Fig{Fig:bMPS} we {show lattice realizations} of $\RP2$ {different from those in the main text}, which results in two different effective conical angles ($2\pi/3$ and $4\pi/3$, respectively) after the cut-and-sew process. Through TN simulations, {we find that the rainbow term changes to $F_{\mathcal{R}} =\frac c 4 \mathcal A\ln \beta$}, where $\mathcal A=7/6$ is a geometric factor. It is then revealed that this logarithmic term can be obtained by substituting two conical angles $\theta =2\pi/3$ and $4\pi/3$ to   {Cardy-Peschel formula} $\frac {c\theta}{24\pi}[(\frac{2\pi}{\theta})^2-1] \ln L$ \cite{Cardy-Peschel-1988S}, which {provides a very intuitive explanation of the rainbow term: they originate from} effective conical singularities.}

\subsection{A. Symmetry and normalization condition of boundary MPS}
\label{App:SymmTN}

\begin{figure}[tbp]
\includegraphics[angle=0,width=0.8\linewidth]{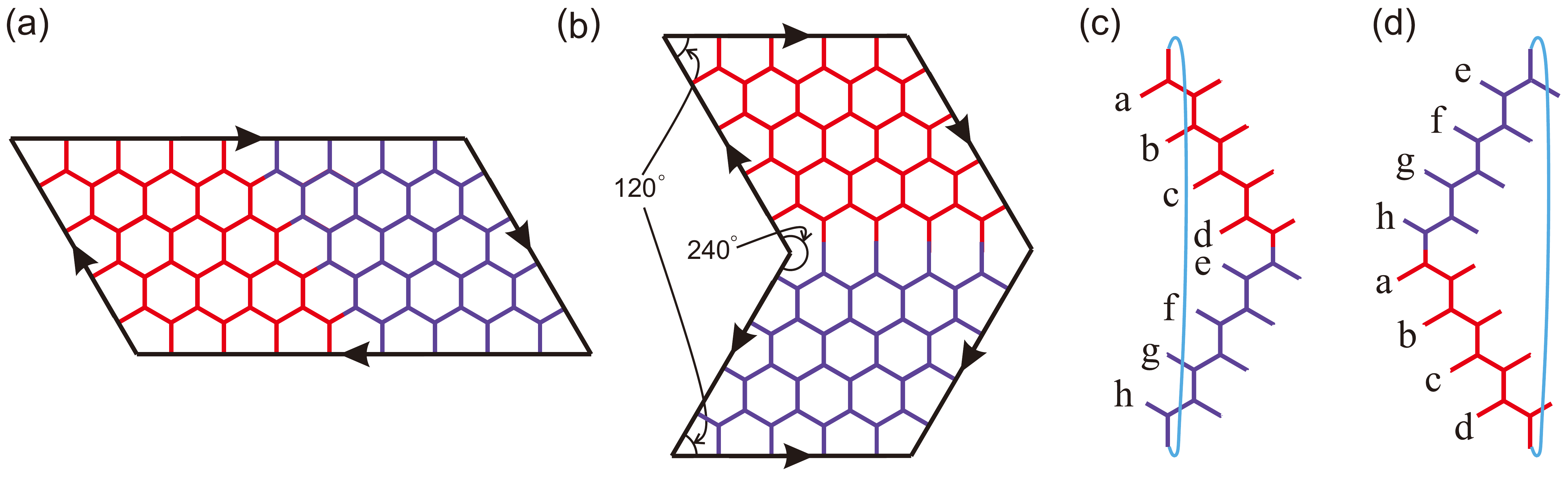}
\caption{(a) A TN realization of $\RP2$ where the edges are glued such that the arrows match. The honeycomb lattice TN is transformed in to (b) with two conical angles, after a cut-and-sew process. The transfer matrix (c) and its transformation (d) by a shift of half the column, which are essentially equivalent, due to vertical PBC. (c) and (d) are also reflection symmetric by construction, implying that the left eigenvector of (c) is identical to the right eigenvector of (d), which is then also equivalent to the right eigenvector of (c), up to a translation of half a column backward).}
\label{Fig:bMPS}
\end{figure}

  {Since the vertical transfer matrix in Figs.~\ref{Fig:bMPS}(c,d) is no longer real symmetric, we firstly address the glide symmetry of transfer matrices, as well as related TN techniques for correctly extracting the universal data.} We demonstrate that, due to the symmetry in the transfer matrix, a proper normalization of bMPS can be done, which is then of importance in the correct extraction of   {universal} ``boundary" terms.

 As shown in Fig. 2 of the main text, the TNs are defined on either the square or honeycomb lattice. The TN representations are constructed symmetrically in the main text. For example, the square TN has reflection symmetry and thus the transfer matrix has the same left and right eigenvectors, i.e., unnormalized $|\tilde{i}_0\rangle=|\tilde{j}_0\rangle$ (left and right eigenvectors, respectively). In this case, we can set the factors as $\mathcal{N}=\langle \tilde{i}_0 | \tilde{i}_0\rangle=\langle \tilde{j}_0|\tilde{j}_0\rangle$ and then normalize the vectors as $| i(j)_0 \rangle = |\tilde{i}(\tilde{j})_0\rangle / \sqrt{\mathcal{N}}$. 
Given the normalized vectors $| i(j)_0 \rangle$, we can extract the crosscap and rainbow free energy terms by calculating $F_{\mathcal{C}} = \ln|\langle\mathcal C|i_0\rangle|$, and $F_{\mathcal{R}} = \ln|\langle\mathcal R|i_0\rangle|$, respectively. 

However, for the construction in \Fig{Fig:bMPS}(c), the transfer matrix is no longer parity symmetric and, at first glance, there seems to be some subtlety in determining universal data associated with each boundary. If one normalizes the left and right eigenvectors of the transfer matrix in a naive manner, i.e., $\langle i_0|i_0\rangle=\langle j_0|j_0\rangle=1$, the mutual normalization condition
\begin{equation}
\langle j_0|i_0\rangle=1
\label{klein-NORM}
\end{equation}
will not be guaranteed in general, and as a result, $\ln{|\langle\mathcal{B_L} | i_0 \rangle|}$ ($\ln{|\langle j_0 |\mathcal{B_R} \rangle|}$) does not produce the desired crosscap (rainbow) term. If one aims at the normalization condition in Eq.~(\ref{klein-NORM}) and there seems existing an arbitrariness to distribution the normalization factor between $|i_0 \rangle$ and $|j_0 \rangle$, which then affects the determination of the crosscap and rainbow terms.

\begin{figure}[htbp]
\centering
\includegraphics[width=.5\textwidth]{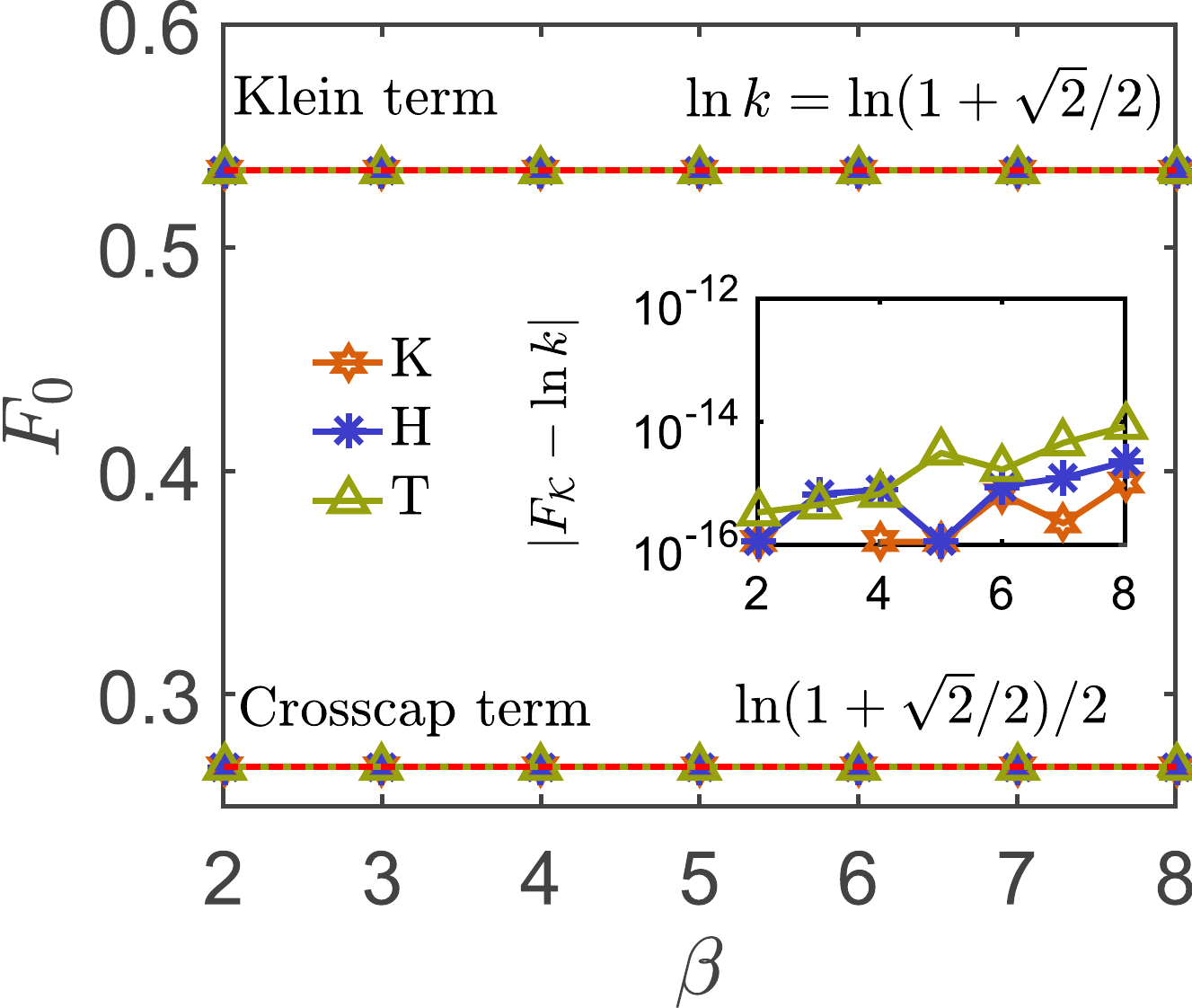}
\caption{  {Crosscap term $F_{\mathcal{C}}$ and Klein term $F_{\mathcal{K}}$ of Ising model on various lattices, the corresponding (honeycomb) TN realization of $\RP2$ is shown in Fig.~\ref{Fig:bMPS}. The retained number of states $D=256$ in the simulations. \label{FigS5}}}
\end{figure}

  {The solution to this difficulty is to note that this arbitrariness can be removed by exploiting the glide symmetry between} $|i_0\rangle$ and $|j_0\rangle$, which enables a proper normalization of both $|i_0\rangle$ and $|j_0\rangle$. As elaborated in \Fig{Fig:bMPS}, unnormalized $|\tilde{j}_0\rangle$ can be related to $|\tilde{i}_0\rangle$ via a shift operation (of half a column), and thus $\langle \tilde i_0 |\tilde i_0 \rangle=\langle \tilde j_0 | \tilde j_0 \rangle$. Therefore, in order to satisfy the condition Eq.~\eqref{klein-NORM}, one can evenly distribute the normalization factor $\mathcal{N}_f=\langle \tilde{j}_0|\tilde{i}_0\rangle$ to $|\tilde{i}_0\rangle$ and $\langle \tilde{j}_0|$, namely $| i_0\rangle =  \tfrac{1}{\sqrt{\mathcal{N}_f}} | \tilde{i}_0  \rangle$ and $\langle j_0| =  \tfrac{1}{\sqrt{\mathcal{N}_f}}\langle \tilde{j_0} |$. After this normalization, it turns out that both $\ln |\langle \mathcal{C} | i_0\rangle|$ and $\ln |\langle j_0 | \mathcal{C} \rangle|$ result in the same crosscap term, i.e., exactly half of the Klein term, $F_{\mathcal{C}}=\frac{1}{2} \ln{k}$, as shown in Fig.~\ref{FigS5}. 
%For transfer matrix of some other TN, one may not extract the crosscap entropy with the specific boundary if the transfer matrix has no symmetry. Usually it's not the case in statistic model.

At last, we have to mention that in the cases above the symmetries (reflection or glide symmetry) in the TN is of key importance in extracting the universal term associated with a specific boundary. In case there is no such kind of symmetries present in the transfer matrix of TN, only the total universal term contributed from both edges can be calculated and it is no longer convenient to distribute it. 

\subsection{B. Crosscap term}
  {Figure \ref{FigS5} shows the crosscap terms and Klein terms of Ising model on hexagonal, kagome and triangular lattices, with corresponding TN representations in \Fig{Fig:bMPS}. It is very surprising that the crosscap/Klein term for all three cases are all extraordinarily accurate, i.e., equal to $F_{\mathcal{C}}=\frac 1 2\ln{(1+\sqrt{2}/2)}$ or $F_{\mathcal{K}}=\ln{(1+\sqrt{2}/2)}$ up to numerical noises/small computational errors. Analytical calculations reveal that even a single row of those three lattices can already produce the universal value exactly, and the analytical solution of single-row kagome lattice can be found in Section IV.D.}

  {The universal crosscap term $\ln k/2$ we found here is very remarkable since it does not depend on the specific geometric details, but only relates to the non-orientable topology. On the Klein bottle we have found $\ln k$ constant entropy \cite{UniEntropy-2017S}, while here on $\RP2$ we discover possibly the smallest unit of the ``twist entropy", i.e., $\frac{1}{2} \ln{k}$.}

\subsection{C. Rainbow free energy term }
\begin{figure}[htbp]
\centering 
\includegraphics[width=.5\textwidth]{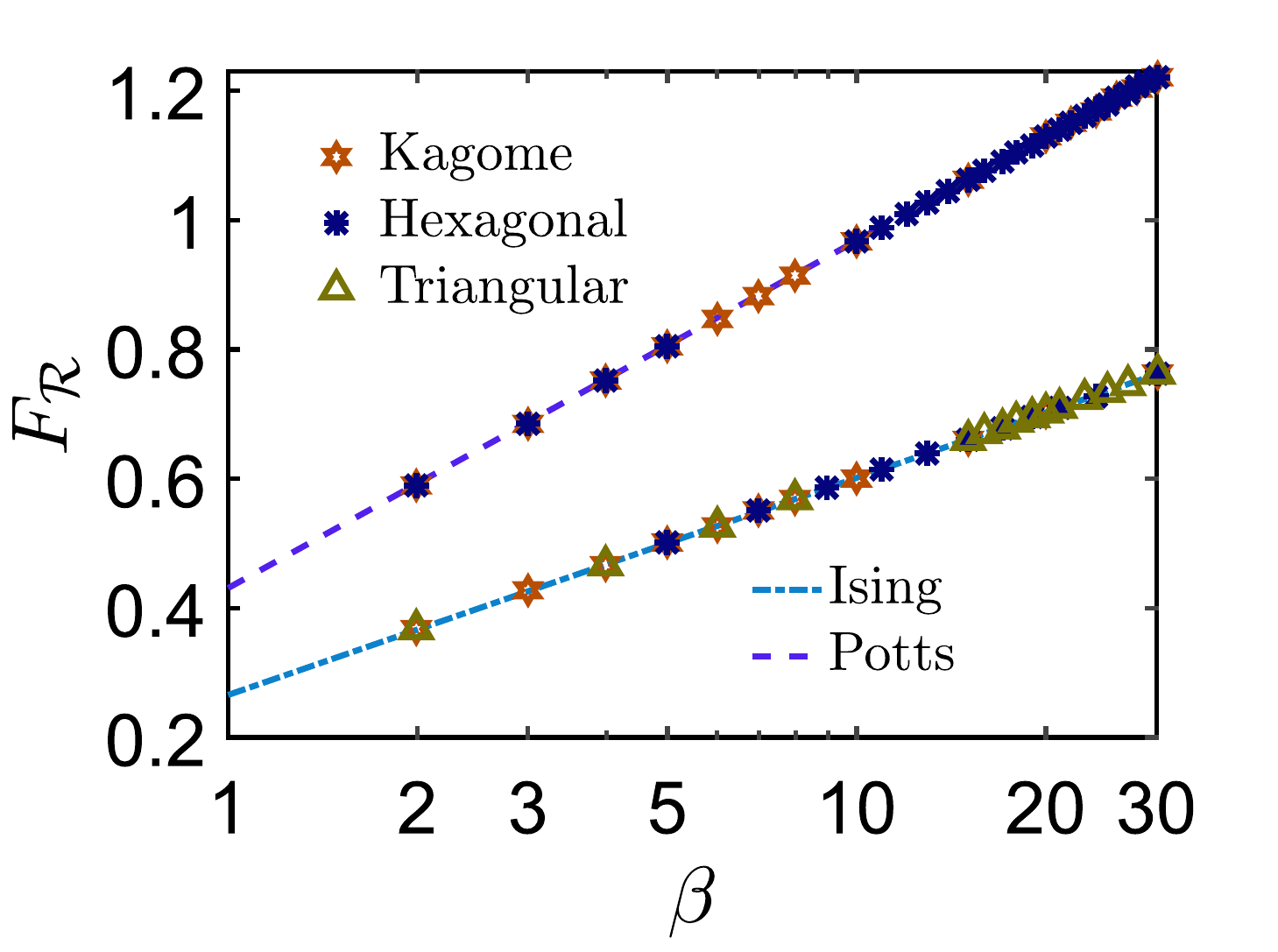}
\caption{TN simulation data with $\chi=500$ for Ising, and $\chi=200$ for Potts model. Dashed lines represent the linear fitting $F_
\mathcal R=\frac c 4\times \frac 7 6 \ln \beta+b_0$, $b_0$ is some non-universal constant\label{FigS6}}
\end{figure}
  {
In the main text, we have shown that with angle $\theta = \pi$ conical singularities,  we can recover the rainbow term $\frac{c}{4} \ln\beta$.
In Fig.~\ref{FigS6}, we show that the rainbow terms in TN with different lattice geometries in \Fig{Fig:bMPS} (but the same non-orientable topology, i.e., $\RP2$). The slopes of the lines in Fig.~\ref{FigS6} is no longer $\frac c 4$ as in the main text, but equal $ \mathcal{A}c/4$, with a geometry factor $\mathcal A \simeq 1.165$. To understand the existence of the geometry factor $\mathcal{A}$, we again associate the logarithmic term to the effective conical singularities, as shown in Fig. \Fig{Fig:bMPS}, with angles $\theta_1=4\pi/3$ and $\theta_2=2\pi/3$ now. Substituting the two angle values to Cardy-Peschel formula
\begin{equation}
F_\theta = \frac {c\theta}{24\pi}[(\frac{2\pi}{\theta})^2-1] \ln L,
\end{equation}
one gets $F_{\theta_1}=\frac{5c}{72}\ln \beta$ and $F_{\theta_2}=\frac{2c}{9}\ln \beta$, and $F_\mathcal R=F_{\theta_1}+F_{\theta_2}=\frac c 4 \frac 7 6 \ln \beta$, by assuming the characteristic system size $L \sim \beta$. Some numerical results are summarized in Tab.~\ref{tab}.

One may also think of the factor $\mathcal{A}$ as $1/\sin \frac{2\pi}{3} \simeq 1.1547005$, which is another possible formula of the geometric factor $\mathcal{A}$, as the rotation symmetry in hexagonal TN could be responsible for it. However, there are two reasons to favor the latter. Firstly, the hexagonal lattice in the main text contribute the logarithmic term with $\mathcal A=1$. If the factor is due to rotational symmetry, it should also be $\mathcal A\neq 1$ there too, which however is not the case. Secondly, there is some appreciable, i.e., $\sim$ 1\% difference between the values  $1/\sin \frac{2\pi}{3} \simeq 1.155$ and $7/6 \simeq 1.1666667$. Our numerical results in Tab.~\ref{tab}, with fine resolution to distinguish the two, clearly support the latter.}

\begin{table*}[htbp]
 % \centering
 \footnotesize
  \caption{Fitting logarithmic terms of hexagon-like lattice (coordination number $z$). The data is fits by $\beta>3$. For Potts model, we also get rid of the data with $\beta >25$ for relatively large errors.}
    \begin{tabular}{|c|c|c|c|c|c|}
    \hline
    \hline
    Model & \multicolumn{3}{c|}{Ising ($c=0.5$)} & \multicolumn{2}{|c|}{Potts (0.8)}    \\ \hline
    Lattice & {H} & {T} & K & H & K  \\ \hline
    $T_c$&  \multicolumn{2}{c|}{$\cosh(2/T_c)=\sec(\pi/z)$ \cite{PhysRev.79.357S}} &$\frac{4}{\ln(3+2\sqrt3)}$ \cite{doi:10.1143-ptp-10.2.158S} & 0.6738 \cite{RevModPhys.54.235S} &1.5849 \cite{RevModPhys.54.235S}  \\ \hline
    Slopes $D$&  0.14577 & 0.14581 & 0.14581 & 0.23296 & 0.232985  \\ \hline
%    $\mathcal{A}$ & $ \sin{\frac{2\pi}{3}}$ & $\sin{\frac{2\pi}{3}}$ & $\sin{\frac{2\pi}{3}}$   & $\sin{\frac{2\pi}{3}}$ & $\sin{\frac{2\pi}{3}}$ \\ \hline
 %      fitted $c$ & 0.502(6) & 0.505(6) & 0.505(6) &  0.802(3) & 0.804(3)  \\ \hline
 	$\mathcal A=4D/c$ &1.16621 &1.16648& 1.16651  & 1.1648&1.1649\\ \hline
    \hline
    \end{tabular}%
  \label{tab}%
\end{table*}%

\subsection{D. Exact Klein Term of Ising model on a Kagome $\Delta-$chain}
\label{App:SYMPO}
\begin{figure}[htbp]
\includegraphics[width=0.7\linewidth]{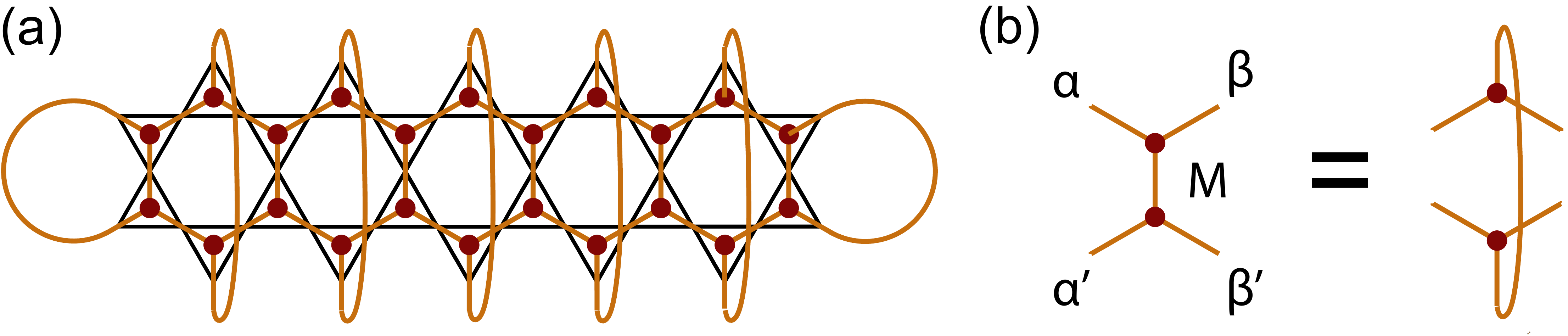}
\caption{TN representation of kagome-lattice Ising model on $\mathbb{K}^2$ with width $\beta=1$. (a) shows a geometry after a cut-and-sew process, which becomes a width-2 lattice with two ``mini-crosscaps" on both ends. (b) depicts the transfer matrix between different sites in (a).
\label{Fig:DChain}}
\end{figure}

For a Klein bottle with width $\beta$=1 [as depicted in \Fig{Fig:DChain}(a), 
also dubbed as the $\Delta$-chains hereafter],
 we can obtain the transfer matrix analytically and thus calculate the residual free energy directly. 
 We take the kagome-lattice Ising model as an example and demonstrate that, on this quasi-1D system, the residual free energy term \textit{exactly} equals $\ln k$, where $k=\sum_\alpha d_\alpha/D$ is the sum of quantum dimensions of CFTs.  

Due to the PBC, the triangles pointing outwards (those on even columns) can be flipped inwards, and the system is essentially of period one. 
The transfer matrix, shown in \Fig{Fig:DChain}(b), is denoted as $M^{\alpha\beta}_{\alpha'\beta'}$. $M$ consists of two rank-three tensors $T_{i,j,k}$,  storing the Boltzmann weight, i.e., $T_{i,j,k}=\exp(-h_{\triangle_{i,j,k}}/T_c)$, where $h_{\triangle_{i,j,k}}=-J(s_is_j+s_js_k+s_ks_i)$ is the Hamiltonian of three spins within the same triangle ($s_i=\pm 1$ labels a Ising spin). By contracting two $T$ tensors, we arrive at
\[
M=
 \begin{pmatrix}
 x^3+x^{-1}&x^1+x^{-1}&x^1+x^{-1}&2x^{-1}\\
 x^1+x^{-1}&2 x^1&2x^{-1}&x^1+x^{-1}\\
 x^1+x^{-1}&2x^{-1}&2 x^1&x^1+x^{-1}\\
 2x^{-1}&x^1+x^{-1}&x^1+x^{-1}&x^3+x^{-1}
 \end{pmatrix}
\]
where $x=\exp(2J/T_c)$, with the basis ordering $(\uparrow \uparrow, \uparrow\downarrow, \downarrow\uparrow, \downarrow\downarrow)$. By diagonalizing the (symmetric) transfer matrix $M$, we obtain the dominant eigenvetor $|i_0\rangle$ as $(1,\sqrt2-1,\sqrt2-1,1)/(2 \sqrt{2-\sqrt{2}})$, at the critical point $T_c/J=\frac{4}{\ln(3+2\sqrt3)}$. We can calculate the residual free energy term according Eq. (1) in the main text, using the crosscap boundary $|\mathcal C\rangle=(1,0,0,1)$. As a result, $F_{\mathcal K}=\ln (1+\sqrt2/2)$, equals the CFT prediction exactly!

This calculation remarkably confirm that the residual term value equals exactly the universal Klein term in such a narrow strip. For Ising model on wider kagome strips ($\beta>1$), we see the residual free energy term stays exactly $\ln (1+\sqrt2/2)$ throughout, in Fig.~\ref{FigS5}. In addition, from Fig.~\ref{FigS5} we observe that Ising models with tilted square and honeycomb TNs also possess such a nice property, i.e., the residual free energy term perfectly coincides with $\ln (1+\sqrt2/2)$ even on lattices with smallest possible width ($\beta=1$).

\end{document}